\documentclass[conference]{IEEEtran}
\IEEEoverridecommandlockouts
\usepackage{cite}
\usepackage{amsmath,amssymb,amsfonts}
\usepackage{algorithmic}
\usepackage{graphicx}
\usepackage{textcomp}
\usepackage{xcolor}
\usepackage{subfigure}

\def\BibTeX{{\rm B\kern-.05em{\sc i\kern-.025em b}\kern-.08em
    T\kern-.1667em\lower.7ex\hbox{E}\kern-.125emX}}

\begin{document}

\title{Debiasing Graph Transfer Learning via Item Semantic Clustering for Cross-Domain Recommendations}

\author{\IEEEauthorblockN{Zhi Li}
\IEEEauthorblockA{
\textit{Osaka University}\\
Osaka, Japan \\
li.zhi@ist.osaka-u.ac.jp}
\and
\IEEEauthorblockN{Daichi Amagata}
\IEEEauthorblockA{
\textit{Osaka University}\\
Osaka, Japan \\
amagata.daichi@ist.osaka-u.ac.jp}
\and
\IEEEauthorblockN{Yihong Zhang}
\IEEEauthorblockA{
\textit{Osaka University}\\
Osaka, Japan \\
yhzhang7@gmail.com}
\and 
\IEEEauthorblockN{Takahiro Hara}
\IEEEauthorblockA{
\textit{Osaka University}\\
Osaka, Japan \\
hara@ist.osaka-u.ac.jp} 
\and 
\IEEEauthorblockN{Shuichiro Haruta}
\IEEEauthorblockA{
\textit{KDDI Research, Inc.}\\
Saitama, Japan \\
sh-haruta@kddi-research.jp}
\and
\IEEEauthorblockN{Kei Yonekawa}
\IEEEauthorblockA{
\textit{KDDI Research, Inc.}\\
Saitama, Japan \\
ke-yonekawa@kddi-research.jp}
\and
\IEEEauthorblockN{Mori Kurokawa}
\IEEEauthorblockA{
\textit{KDDI Research, Inc.}\\
Saitama, Japan \\
mo-kurokawa@kddi-research.jp}
}

\maketitle

\begin{abstract}
Deep learning-based recommender systems may lead to over-fitting when lacking training interaction data.
This over-fitting significantly degrades recommendation performances.
To address this data sparsity problem, cross-domain recommender systems (CDRSs) exploit the data from an auxiliary source domain to facilitate the recommendation on the sparse target domain. 
Most existing CDRSs rely on overlapping users or items to connect domains and transfer knowledge.
However, matching users is an arduous task and may involve privacy issues when data comes from different companies, resulting in a limited application for the above CDRSs.
Some studies develop CDRSs that require no overlapping users and items by transferring learned user interaction patterns. 
However, they ignore the bias in user interaction patterns between domains and hence suffer from an inferior performance compared with single-domain recommender systems. 
In this paper, based on the above findings, we propose a novel CDRS, namely semantic clustering enhanced debiasing graph neural recommender system (SCDGN), that requires no overlapping users and items and can handle the domain bias.
More precisely, SCDGN semantically clusters items from both domains and constructs a cross-domain bipartite graph generated from item clusters and users.
Then, the knowledge is transferred via this cross-domain user-cluster graph from source to the target.
Furthermore, we design a debiasing graph convolutional layer for SCDGN to extract unbiased structural knowledge from the cross-domain user-cluster graph. 
Our Experimental results on three public datasets and a pair of proprietary datasets verify the effectiveness of SCDGN over state-of-the-art models in terms of cross-domain recommendations.
\end{abstract}


\section{Introduction}
As an effective technique to overcome information overload, recommender systems (RSs) can filter out items (e.g., products on Amazon and movies on Netflix) that users may like and provide personal recommendations for users to improve their experiences and to generate commercial profits. 
RSs predict users' future item-interacted behaviors, e.g., purchasing in e-commerce and rating in the case of online movies, where the predictions are inferred by models learned from their past interaction behaviors \cite{ma2008pmf,hsieh2017collaborative,li2022trends,amagata2019dynamic}. 
Deep learning has been employed in RSs, with its well-designed structures and the large number of learnable parameters, to better model users' complex interaction patterns \cite{nguyen2020transferability}.
In real-world services, most users interact with only a few items, particularly in start-up companies and when companies develop new services. 
In such scenarios, deep learning-based RSs may lead to over-fitting because of the sparse interaction data \cite{trong2018infipf,li2022hml4rec,wang2020dnn,wang2019preliminary}, which significantly degrades the recommendation performance.

\subsection{Motivation}
To address the aforementioned data sparsity problem, cross-domain recommender systems (CDRSs) have been developed. 
CDRSs exploit the data from an auxiliary domain (i.e., a source domain) to facilitate the inference process in a target domain. 
Existing CDRSs are categorized into two approaches: multi-task learning and transfer learning \cite{Zhu2021CDRCPP,yonekawa2019heterogeneous,kurokawa2018virtual}.
In multi-task learning-based CDRSs, some neural layers or user (item) embedding are shared between domains \cite{zhu2019dtcdr,zhu2020dtcdr,li2022RecGURU}. 
These shared layers or embedding are optimized by fitting the recommendation tasks to both source and target domains.
As a result, the trained layers or embedding can learn the knowledge from both domains, and hence can provide more accurate recommendations than single-domain recommendations. 
In contrast, transfer learning-based CDRSs focus on recommendations in the target domain \cite{liu2020transgcf,Guo2021DAGCN,liu2022CFAA}.
These CDRSs extract knowledge from the source domain and use the learned knowledge to improve the target recommendations.

Most existing CDRSs, including the above ones, bridge domains by matching user information and transferring knowledge from the source domain to the target domain via overlapping users. 
These overlapping users are, however, not always available in real-world services \cite{lyu2019behavior}.
In addition, user matching is an arduous task and may involve privacy issues, particularly when data come from different companies. 
Some CDRSs, such as MMT-DRR \cite{Krishnan2020mtt}, RecSys-DAN \cite{wang2020rsdan}, and CFAA \cite{liu2022CFAA}, therefore extract user interaction patterns from the source domain and transfer the learned interaction patterns to the target domain, where the interaction patterns are defined as learned user (item) embedding or the distribution of predictions.
They define domains as different categories in a Web service, such as Amazon Book and Amazon Movie, or different places in a real-world service, such as the restaurant visit records in different cities.
By doing so, they can merge the source and target domains with domain-shared side information, such as user profiles and item contents.  
Unfortunately, user interaction patterns have been observed to be strongly domain-dependent, particularly when these domains come from different services \cite{li2021debias}.
These CDRSs hence suffer from the domain bias in user interaction patterns.

\subsection{Contribution}
From the above observations, it can be seen that a technique which demands no user matching and can alleviate the bias of interaction patterns is required.
Motivated by this, we develop a new CDRS, namely semantic clustering enhanced debiasing graph neural recommender system (SCDGN), that is applicable to different services sharing no entities, including users and items.
To achieve this, we generate a cross-domain user-cluster graph to bridge two domains, where the graph consists of users and item clusters. Since item clusters are generated from textual information on items (e.g. movie titles and web page contents), user-cluster graph can merge the semantic interaction information of different domains and user matching becomes unnecessary.  
Then, We extract knowledge from both target item-level and cross-domain cluster-level interaction graphs by devising a CDRS variant of LightGCN \cite{he2020lightgcn}.
This variant is inspired by the success of graph convolutional networks \cite{hamilton2017gcn} in extracting complex high-hop neighbor information from graph structures.
To handle the bias of user interaction patterns in different domains, we develop a novel debiasing graph convolutional layer to learn unbiased knowledge from the cross-domain user-cluster graph.
More precisely, we design adaptive debiasing vectors for users and item clusters to weight edges in the cross-domain user-cluster graph. 
Moreover, inspired by the effectiveness of debiasing learning \cite{li2021debias}, we develop two-level restrictions to learn the above debiasing vectors. 

To summarize, this work makes the following main contributions:
\begin{itemize}
\item   We propose a novel semantic cluster-based domain merge approach to make the interaction information transferable at an item-cluster level. By doing so, Our CDRS does not require user matching.
\item   We develop a debiasing cluster-enhanced cross-domain graph convolutional model to transfer knowledge and alleviate the bias in interaction patterns between domains. 
\item   We conduct extensive experiments on three public datasets and a pair of proprietary dataset to evaluate the effectiveness of our CDRS. The results demonstrate that our proposal outperforms state-of-the-art methods.
\end{itemize}
This is a full version of \cite{li2022debiasing}.

\section{Related Work \label{sec:related}}
\subsection{Cross-domain Recommender Systems \label{sec:CDR}}
To mitigate the data sparsity problem, CDRSs leverage data from an auxiliary source domain to facilitate recommendations in the sparse target domain.
Some existing CDRSs require overlapping users to bridge the source and the target domains \cite{zhang2020itemsetmap,li2022RecGURU,li2021doml,cao2022disencdr}.
With these users, the source domain can transfer individual-level knowledge to the target domain. 
Learning transformation function and domain adversarial learning are two promising directions for cross-domain recommendations. 
The former one learns transformation functions to transfer user or item embeddings from the source domain to the target domain.
For example, CGN \cite{zhang2020itemsetmap} proposed a novel generative adversarial network to transfer item embeddings in a set manner. 
DOML \cite{li2021doml} learned a latent orthogonal metric mapping to transfer the user embedding between domains.   
PTUPCDR \cite{zhu2022ptupcdr} considered the bias caused by personal differences and introduced a meta-learning method that learns user-specific transformation functions to handle the personal difference bias for cross-domain recommendations.
Meanwhile, domain adversarial learning aligns source and target embedding spaces to transfer knowledge from the source domain to the target domain. 
For example, DARec \cite{yuan2019darec} developed a deep domain adaptation model to transfer rating patterns. 
RecGURU \cite{li2022RecGURU} introduced a transformer network and minimized Kullback–Leibler divergence between the learned distributions of latent user
representations to learn the domain-invariant embeddings for cross-domain sequential recommendations. 

However, matching users is an arduous task and may involve privacy issues in most real-world applications.
Considering privacy and the scalability of methods, some studies avoid user alignment and transfer distribution-level knowledge from the source domain to the target domain.
MMT-DRR \cite{Krishnan2020mtt} regularized the target domain’s user and item embedding space with the embedding space learned in source domains.
However, MMT-DRR cannot work without domain-shared contextual information.
ESAM \cite{Zhihong2020ESAM} and CFAA \cite{liu2022CFAA} removed the requirement of domain-shared contextual information and aligned the attribution distribution and correlation between source and target embedding spaces to transfer knowledge.
Besides, RecSys-DAN \cite{wang2020rsdan} proposed a novel discriminator and minimized the divergence of the predictions between the source domain and the target domain for knowledge transformation.
Unfortunately, bias in interaction patterns between domains may degrade the recommendation performances of CDRSs. The methods mentioned above do not consider this domain bias issue.
We hence propose a CDRS to alleviate this bias without using overlapping users.

\subsection{Graph Convolution in Recommendations \label{sec:GNN-RS}}
Recently, graph neural networks (GNNs) have been employed in RSs to guide the embedding learning by exploiting user-item graph structures \cite{wu2021SGL,xia2021GMN,chen2021MSGCN,wang2019kgat}. 
PinSage \cite{ying2018PinSage} and NGCF \cite{wang2019ngcf} defined the information propagation as aggregation of the embeddings of neighbors to enhance the target node's (i.e., users' or items') embedding.
Considering that recommender systems often use one-hot embedding (i.e., less information than images and text), SGCN \cite{wu2019SGCN} and LightGCN \cite{he2020lightgcn} further simplified and customized graph models to avoid over-fitting. 
In addition, some GNN-based CDRs also alleviate the sparse problem by combining the complex high-order graph structural information from the source to the target \cite{Guo2021DAGCN,xu2021recdr,zhu2020dtcdr,liu2020transgcf}. 
For example, GA-DTCDR \cite{zhu2020dtcdr} constructed heterogeneous graphs to learn user and item embeddings and developed an element-wise attention mechanism to combine the embeddings of users learned from both domains. 
However, the above-mentioned GNN-based CDRs require overlapping users to connect domains and ignore the domain bias in user preferences patterns. 
BiTGCF \cite{liu2020transgcf} developed a domain-specific feature propagation layer to handle the domain bias, but it still requires overlapping users to fuse domain information.
In light of the above causal view, we develop a GNN-based CDRS that requires no overlapping users and can handle the domain bias. 

\begin{figure*}[!t]
    \centering
    \includegraphics[width=1.0\linewidth]{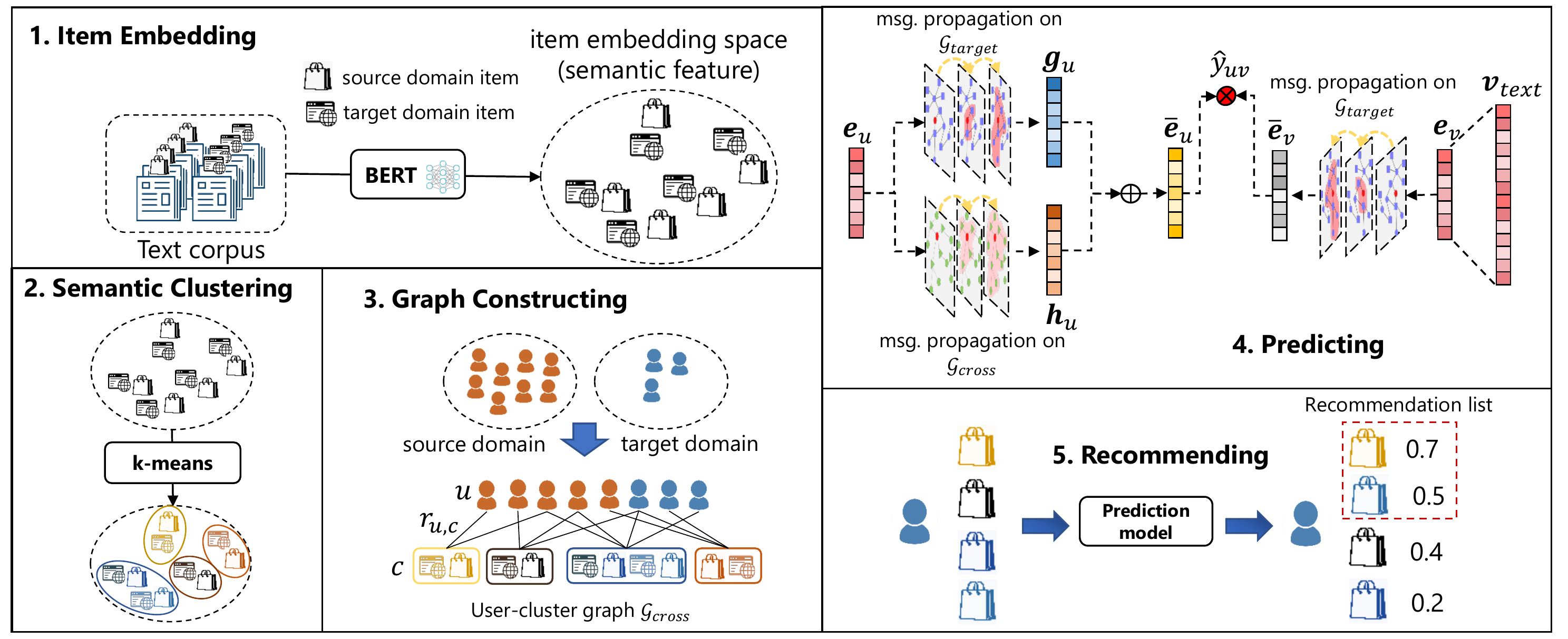}
    \caption{
    Overview of our proposed CDRS.
    (1) The item embedding learns semantic embeddings for all items in source and target domains.
    (2) All items in the source and target domains are then clustered by their semantic embeddings.
    (3) A cross-domain user-cluster graph is constructed to merge the two domains' interaction information at a semantic cluster level.
    (4) A debiasing graph convolutional neural network makes predictions by leveraging the interaction information from the target user-item and cross-domain user-cluster graphs. 
    (5) Finally, recommendations are produced by the prediction results.}
    \label{fig:frameOverview}
\end{figure*}

\section{Preliminaries \label{sec:Preli}}
\subsection{Problem Formulation \label{sec: problem}}
In this work, we define the top-K recommendations in a sparse domain as our recommendation task.
We consider an auxiliary source domain $\mathcal{D}_s$ and a sparse target domain $\mathcal{D}_t$. 
$\mathcal{D}_t$ contains $\mathcal{U}_t$, $\mathcal{V}_t$, and $\mathcal{R}_t$, where $\mathcal{U}_t$ ($\mathcal{V}_t$) denotes the user (item) set and $\mathcal{R}_t$ is the interaction set between them.
Similarly, $\mathcal{D}_s$ contains $\mathcal{U}_s$, $\mathcal{V}_s$, and $\mathcal{R}_s$.
There is no overlap between the user and item sets of $\mathcal{D}_s$ and $\mathcal{D}_t$.

To address the data sparsity problem, we consider the semantic clustering information of items $\mathcal{C}$ extracted from both the source and target domains, because this enhances the sparse interactions in the target domain.
As a result, each interaction $r \in \mathcal{R}_t$ is a tuple $r = (u, v, c)$, where $u \in \mathcal{U}_t$, $v \in \mathcal{V}_t$, and $c \in \mathcal{C}$.
For each user $u \in \mathcal{U}_t$, we predict a preference score $\hat{y}_{u,v}$ for each item $v \in \mathcal{\bar{V}}_u = \{v \in \mathcal{V}_t, v \notin \mathcal{V}_u\}$, where $\mathcal{V}_u$ is the items that interacted with $u$.
We then rank the items in $\mathcal{\bar{V}}_u$ according to their preference scores and recommend the top-K items with the largest scores to $u$. 

\subsection{Simplified Graph Convolution for RS \label{sec: lightgcn}}
LightGCN \cite{he2020lightgcn} is a graph convolution network that refines user and item embedding by extracting structural information, particularly high-hop neighbors, from the user-item interaction graph.
ID-based embedding in RSs contains less available information than words in text and pixels in images.
Hence, LightGCN removes the non-linear projection and the self-connection operations from its massage propagation.   
More precisely, the $l$-th simplified graph convolution (i.e., massage propagation) layer in LightGCN is defined as:
\begin{equation}
\begin{aligned}
&\mathbf{e}_{u}^{(l+1)}=\sum_{v \in \mathcal{V}_u} \frac{1}{\sqrt{\left|\mathcal{V}_u\right|} \sqrt{\left|\mathcal{U}_v\right|}} \mathbf{e}_{v}^{(l)}, \\
&\mathbf{e}_{v}^{(l+1)}=\sum_{u \in \mathcal{U}_v} \frac{1}{\sqrt{\left|\mathcal{U}_v\right|} \sqrt{\left|\mathcal{V}_u\right|}} \mathbf{e}_{u}^{(l)},
\end{aligned}
\end{equation}
where $\mathbf{e}_{u}^{(0)}$ and $\mathbf{e}_{v}^{(0)}$ are the ID embeddings of user $u$ and item $v$, respectively.
$\mathcal{U}_v$ is a set of users that interacted with $v$.
This graph convolutional layer has been analytically and empirically proven to be effective in accelerating the training process and alleviating the data sparsity problem.

\section{Proposed Method \label{sec:propose}}
Motivated by the observations that existing CDRSs require user matching or suffer from domain-dependent interaction patterns, we propose a novel CDRS, which is depicted in Figure \ref{fig:frameOverview} and does not have these drawbacks.
The numbers below correspond to the ones in Figure \ref{fig:frameOverview}.

(1) Our idea for avoiding user matching is to merge the source and target domains by using \textit{semantic information of items}.
Such semantic information can be extracted via a pre-trained representation extractor BERT \cite{devlin2019Bert}.
(2) Then, we cluster all items in the source and target domains via their semantic information.
(3) After that, from a graph structure that has edges between users and corresponding item clusters in the semantic space, we can obtain \textit{merged} cluster-level interaction patterns of the two domains.
(4) Our prediction model fuses the user embeddings enhanced via the target user-item and the cross-domain user-cluster interaction graphs to improve the expression of users with the semantic cluster knowledge from the source domain.
In particular, we develop a novel debiasing graph convolutional layer to alleviate domain bias and extract unbiased structural knowledge.
(5) Finally, thanks to the above novel ideas, we can expect a high inner product of a user and an item to which she would prefer, thus yielding in an accurate recommendation list \cite{hirata2022cardinality,hirata2022solving,amagata2021reverse,nakama2021approximate}.
Table \ref{tab:notation} summarizes important notations used in this paper.

\begin{table}[t]
    \begin{center}
        \caption{Summary of notations}
        \label{tab:notation}
        \begin{tabular}{l|l}    
                \hline
                Notation &Description                    \\ 
                \hline
                $u$& a user \\
                $v$& an item \\
                $c$& a cluster (a set of items) \\
				$\mathcal{G}_{cross}$& cross-domain user-cluster graph \\
				$\mathcal{G}_{target}$& target user-item graph \\
				$\mathcal{N}_{u}$& neighbor cluster set of $u$ in $\mathcal{G}_{cross}$ \\
				$\mathcal{N}_{c}$& neighbor user set of $c$ in $\mathcal{G}_{cross}$ \\
				$\mathcal{M}_{u}$& neighbor item set of $u$ in $\mathcal{G}_{target}$ \\
				$\mathcal{M}_{v}$& neighbor user set of $v$ in $\mathcal{G}_{target}$ \\
				$\mathbf{v}_{txt}$, $\mathbf{c}_{txt}$& semantic vectors of $v$ and $c$  \\
				$\mathbf{e}_{u}$, $\mathbf{e}_{v}$, $\mathbf{e}_{c}$& embedding of $u$, $v$, and $c$ \\
				$\mathbf{a}_u$, $\mathbf{a}_c$& debiasing vectors of $u$ and $c$ \\
				$\mathbf{\bar{e}}_{u}$, $\mathbf{\bar{e}}_{v}$, $\mathbf{\bar{e}}_{c}$& unbiased finel embedding of $u$, $v$, and $c$ \\
				$\mathbf{\bar{e}}^{\prime}_{u}$, $\mathbf{\bar{e}}^{\prime}_{c}$& biased finel embedding of $u$ and $c$ \\

				$\mathbf{h}_{u}^{(l)}$, $\mathbf{h}_{v}^{(l)}$& $\mathcal{G}_{target}$'s $l$-th graph conv. layer outputs \\
				$\mathbf{g}_{u}^{(l)}$, $\mathbf{g}_{c}^{(l)}$& $\mathcal{G}_{cross}$'s $l$-th debiasing graph conv. layer outputs \\
				$\mathbf{g}^{\prime (l)}_{u}$, $\mathbf{g}^{\prime (l)}_{v}$& $\mathcal{G}_{cross}$'s $l$-th graph conv. layer outputs \\
                \hline
        \end{tabular}
    \end{center}
\end{table}

\subsection{Semantic Domain Fusing \label{sec:embed}}
To semantically fuse the source and target domains, we first embed all items in the source and target domains into a domain-shared embedding space.
Then, we cluster items based on this embedding space and construct a cross-domain user-cluster graph to enhance the interaction information.

\subsubsection{Semantic item embedding} 
Given textual information on items, such as a description of a product, we extract semantic features from the text information to represent items in the source and target domains. 
We apply the token embeddings from a pre-trained BERT \cite{devlin2019Bert} to represent tokens in item text, because this model is learned by sufficient Wikipedia data and hence contains semantic information.
The text of item $v$ is denoted by $\operatorname{text}(v)$, and an semantic embedding of item $v$ is obtained by
\begin{equation}
\begin{aligned}
\mathbf{v}_{txt}= \sum_{w \in \operatorname{text}(v)} \phi_{tf\mbox{-}idf}(w) \cdot \phi_{\mathrm{BERT}}(w),
\end{aligned}
\end{equation}
where $\phi_\mathrm{BERT}(w)$ and $\phi_{tf\mbox{-}idf}(w)$ are respectively the embedding and the tf-idf score of token $w \in \operatorname{text}(v)$.
Note that $\phi_{tf\mbox{-}idf}(w)$ is calculated based on the text corpus collected from both domains. 

\subsubsection{User-cluster graph construction}
We next construct a user-cluster graph.
This aims at merging the source and target domains \textit{without} user, item, and side information alignments.
In addition, high-hop neighbors in this user-cluster graph can yield useful knowledge to improve the recommendation accuracy \cite{he2020lightgcn}.
In Section \ref{sec:predictor}, we leverage this observation through modeling such structures from this graph, which also motivates building this user-cluster graph.

To construct the user-cluster graph, we first run the semantic clustering in Figure \ref{fig:frameOverview}, that is, we cluster all items from the source and target domains in the learned semantic embedding space.
We employ the empirically effective $k$-means clustering \cite{macqueen1967some} method and leave the discussion of more clustering methods as a future work. 
After that, we construct the cross-domain user-cluster graph $\mathcal{G}_{cross}=\{(u,r_{u,c},c)|u\in \mathcal{U},c \in \mathcal{C}\}$ to merge the two domains' semantic-level interaction information, where $\mathcal{U} = \{\mathcal{U}_s, \mathcal{U}_t\}$ and $\mathcal{C}$ respectively denote the user and cluster sets.
The link $r_{u,c} = 1$ indicates that there is an interaction between $u$ and any item belonging to $c$; otherwise $r_{u,c}=0$.

\subsection{Debiasing Graph Convolutional Predictor} \label{sec:predictor}
We here develop a cluster-enhanced debiasing graph convolutional model for recommendations in the sparse target domain. 
Different from existing CDRSs that transfer item interaction patterns directly, our model transfers the semantic clustering interaction patterns via the cross-domain user-cluster graph $\mathcal{G}_{cross}$.
To achieve this, our model fuses $\mathcal{G}_{cross}$ and the target user-item graph $\mathcal{G}_{target}$ to refines the user and item embeddings with structural knowledge from graphs, where $\mathcal{G}_{target}=\{(u,r_{u,v},v)|u\in \mathcal{U}_t, v \in \mathcal{V}_t, r_{u,v} \in \mathcal{R}_t\}$.
This model consists of three main components:
(i) an embedding layer, which learns latent vectors for users and items,
(ii) debiasing graph convolutional layers, which recursively propagate unbiased high-hop neighbor information to refine the user and item vectors, and 
(iii) a prediction layer, which aggregates the user and item representations from all propagation layers and outputs the predictions.

\subsubsection{Embedding layer}
To alleviate the data sparsity problem, we propose a novel approach that projects users into the item embedding space learned in Section \ref{sec:embed}.
Furthermore, we design a metric-invariant dimension reduction approach to control the scale of parameters according to the difficulty of the recommendation task and the sparsity of the training data.  
The item embedding is calculated by a dimension compression layer: $\mathbf{e}_v = \mathbf{W}\mathbf{v}_{txt} + \mathbf{b}$, where $\mathbf{W}$ and $\mathbf{b}$ are the parameters of this layer. 
The dimension of $\mathbf{e}_v$ is much smaller than that of $\mathbf{v}_{txt}$, in order to adapt to the sparse target domain.
The cluster embedding is computed by the same layer: $\mathbf{e}_c = \mathbf{W}\mathbf{c}_{txt} + \mathbf{b}$, where $\mathbf{c}_{txt}$ is the semantic embedding of cluster $c$.
$\mathbf{c}_{txt}$ is defined as the mean pooling of all item semantic embeddings in this cluster and formulated by
\begin{equation}
\begin{aligned}
\mathbf{c}_{txt}= \frac{1}{|\mathcal{V}_c|} \sum_{v \in \mathcal{V}_c} \mathbf{v}_{txt},
\end{aligned}
\end{equation}
where $\mathcal{V}_c$ is the item set in cluster $c$.
The user embedding is defined as the ID embedding $\mathbf{e}_u$, which has the same dimension as that of $\mathbf{e}_v$.  
We measure the cosine similarities $S_{c}$ between items and clusters and minimize the mean squared error of the cosine similarities calculated before and after dimension reduction to ensure the metric invariance, where the error is defined as  
\begin{equation}
\begin{aligned}
\mathcal{L}_{dr} = \frac{1}{|\mathcal{R}_t|}\sum_{(u,v,c) \in \mathcal{R}_t} &\left(\left(S_{c}(\mathbf{e}_v, \mathbf{e}_{v^{-}}) \mbox{-} S_{c}(\mathbf{v}_{txt}, \mathbf{v}_{txt}^{-})\right)^{2}\right.\\
				                                 +  &\left.\left(S_{c}(\mathbf{e}_c, \mathbf{e}_{c^{-}}) \mbox{-} S_{c}(\mathbf{c}_{txt}, \mathbf{c}_{txt}^{-})\right)^{2}\right).
\end{aligned}
\end{equation}
In this equation, $v^{-}$ is a negative item randomly sampled from $\mathcal{\bar{V}}_u$ and $c^{-}$ is the cluster to which $v^{-}$ belongs.
This approach adjusts the embedding dimension and maintains a consistent spatial relationship with the original embedding space.

\subsubsection{Debiasing graph convolutional layers}
Because of the superiority of graph convolutional networks in capturing and modeling structural information from graphs, we develop graph convolutional modules for extracting structural information from the target user-item graph $\mathcal{G}_{target}$ and the cross-domain user-cluster graph $\mathcal{G}_{cross}$.  
More precisely, we employ the state-of-the-art ``light graph convolution'' layer \cite{he2020lightgcn} to propagate graph information because of its effectiveness in alleviating overfitting for our sparse target domain. 
To identify the domain bias in user preference patterns and extract unbiased knowledge from $\mathcal{G}_{cross}$, we propose a novel debiasing graph convolutional layer. 
For each user $u \in \mathcal{U}$, we set an adaptive debiasing vector $\mathbf{a}_u$ to represent her individual domain bias.
For each cluster $c \in \mathcal{C}$, we also set an adaptive debiasing vector $\mathbf{a}_c$.
By doing so, the debiasing factor of user-cluster interaction $r_{u,c}$ can be defined as: $a_{uc} = \mathbf{a}_u^{T}\mathbf{a}_c$. 
The $l$-th debiasing graph convolutional layer for $\mathcal{G}_{cross}$ is formulated as: 
\begin{equation}
\begin{aligned}
&\mathbf{g}_{u}^{(l+1)}=\sum_{c \in \mathcal{N}_u} \frac{1}{\sqrt{\left|\mathcal{N}_u\right|} \sqrt{\left|\mathcal{N}_c\right|}} a_{uc} \cdot \mathbf{g}_{c}^{(l)}, \\
&\mathbf{g}_{c}^{(l+1)}=\sum_{u \in \mathcal{N}_c} \frac{1}{\sqrt{\left|\mathcal{N}_c\right|} \sqrt{\left|\mathcal{N}_u\right|}} a_{uc} \cdot \mathbf{g}_{u}^{(l)},
\label{eq:DGClayer}
\end{aligned}
\end{equation}
where $\mathcal{N}_u = \{c|r_{u,c} = 1, r_{u,c} \in \mathcal{G}_{cross}\}$ is the neighbor cluster set of user $u$ and $\mathcal{N}_c$ is the neighbor user set of cluster $c$.
We define $\mathbf{g}_{u}^{(0)} = \mathbf{e}_{u}$ and $\mathbf{g}_{c}^{(0)} = \mathbf{e}_{c}$.
It is worth mentioning that we detach the gradient computation of the debiasing vectors $\mathbf{a}_u$ and $\mathbf{a}_c$ here for computational efficiency. 
The learning of $\mathbf{a}_u$ and $\mathbf{a}_c$ is left to the proposed restrictions in Section \ref{sec:restrict}.  
For $\mathcal{G}_{target}$, we adopt the standard ``light graph convolution'' layer \cite{he2020lightgcn}, where the $l$-th graph convolutional layer is formulated as:
\begin{equation}
\begin{aligned}
&\mathbf{h}_{u}^{(l+1)}=\sum_{v \in \mathcal{M}_u} \frac{1}{\sqrt{\left|\mathcal{M}_u\right|} \sqrt{\left|\mathcal{M}_v\right|}} \mathbf{h}_{v}^{(l)}, \\
&\mathbf{h}_{v}^{(l+1)}=\sum_{u \in \mathcal{M}_v} \frac{1}{\sqrt{\left|\mathcal{M}_v\right|} \sqrt{\left|\mathcal{M}_u\right|}} \mathbf{h}_{u}^{(l)}.
\end{aligned}
\end{equation}
$\mathcal{M}_u = \{v|r_{u,v} = 1, r_{u,v} \in \mathcal{G}_{target}\}$ is the neighbor item set of user $u$ and $\mathcal{M}_v$ is the neighbor user set of item $v$.
Similarly, we define $\mathbf{h}_{u}^{(0)} = \mathbf{e}_{u}$ and $\mathbf{h}_{v}^{(0)} = \mathbf{e}_{v}$.

\subsubsection{Prediction layer}
We next refine $\mathbf{e}_u$, $\mathbf{e}_v$, and $\mathbf{e}_c$ by using the extracted graph structure information.
The final representation is produced by aggregating the embeddings obtained at each graph convolutional layer:
\begin{equation}
\begin{aligned}
&\mathbf{\bar{e}}_{u}=\sum_{l=0}^{P} \mathbf{g}_{u}^{(l)} + \sum_{l=0}^{Q} \mathbf{h}_{u}^{(l)},  \\
&\mathbf{\bar{e}}_{v}=\sum_{l=0}^{Q} \mathbf{h}_{v}^{(l)}; \quad \mathbf{\bar{e}}_{c}=\sum_{l=0}^{P} \mathbf{g}_{c}^{(l)},
\end{aligned}
\end{equation}
where $P$ and $Q$ are the numbers of debiasing graph convolutional layers for $\mathcal{G}_{cross}$ and graph convolutional layers for $\mathcal{G}_{target}$, respectively.
It is worth mentioning that $\mathbf{e}_{u}$ is refined by the structural information from both the target and cross-domain graphs, i.e., the knowledge from both the item and cluster level interactions.

Finally, the preference score is defined as the inner product of the user and item final representation:
\begin{equation}
\begin{aligned}
\hat{y}_{u v}=\mathbf{\bar{e}}_{u}^{T} \mathbf{\bar{e}}_{v}
\end{aligned}
\end{equation}

\subsection{Restrictions for Debiasing Learning} \label{sec:restrict}   
The previous debiasing learning \cite{li2021debias} calculates their restrictions via overlapping users and domain-shared item attributions, e.g., category, seller, brand, and price, resulting in a limited application.  
Besides, it directly sets adaptive debiasing factors for each user-item interaction and optimizes them separately.
In other words, the learning of a debiasing factor only relies on the corresponding interaction and thus suffers from a severe overfitting issue.
Based on these findings, we get hints from the matrix factorization algorithm and re-define the debiasing factor $a_{uc}$ as the inner product of the corresponding user debiasing vector $\mathbf{a}_u$ and cluster debiasing vector $\mathbf{a}_c$. 
Our approach learns $\mathbf{a}_u$ and $\mathbf{a}_c$ via the restriction losses at both prediction and individual levels. 

\subsubsection{Restriction in prediction level}
As a debiasing factor, $a_{uc}$ is demanded to produce unbiased prediction $\hat{y}_{u c}$ from the biased version $\hat{y}^{\prime}_{u c}$.
To achieve this, we set a restriction loss $\mathcal{L}_{rsp}$ that measures the mean squared error between $\hat{y}_{u c}$ and $a_{uc} \cdot \hat{y}^{\prime}_{u c}$.
$\mathcal{L}_{rsp}$ is defined as:
\begin{equation}
\begin{aligned}
\mathcal{L}_{rsp}= \frac{1}{|\mathcal{R}_t|}\sum_{(u,v,c) \in \mathcal{R}_t} &\left(\hat{y}_{u c} - a_{u c} \cdot \hat{y}^{\prime}_{u c}\right)^{2},
\end{aligned}
\end{equation}   
where $\hat{y}_{u c}=\mathbf{\bar{e}}_{u}^{T} \mathbf{\bar{e}}_{c}$ and $\hat{y}^{\prime}_{u c}=\mathbf{\bar{e}}^{\prime T}_{u} \mathbf{\bar{e}}^{\prime}_{c}$. 
The biased user embedding $\mathbf{\bar{e}}^{\prime}_{u}$ and the biased cluster embedding $\mathbf{\bar{e}}^{\prime}_{c}$ aggregate the output of every graph convolutional layers and are formulated as:
\begin{equation}
\begin{aligned}
\mathbf{\bar{e}}^{\prime}_{u}=\sum_{l=0}^{P} \mathbf{g}^{\prime (l)}_{u}; \quad
\mathbf{\bar{e}}^{\prime}_{c}=\sum_{l=0}^{P} \mathbf{g}^{\prime (l)}_{c},
\end{aligned}
\end{equation}
where $\mathbf{g}^{\prime (l)}_{u}$ and $\mathbf{g}^{\prime (l)}_{u}$ are the user and cluster aggregation result of the $l$-th graph convolutional layer that can be computed by Equation \ref{eq:DGClayer} without the debasing factor.
By minimizing $\mathcal{L}_{rsp}$, we can ensure a consistent result between the unbiased prediction and the prediction produced by the debiasing graph convolutional layers. 
As a result, $\mathcal{L}_{rsp}$ constrains the embedding space of $\mathbf{a}_u$ and $\mathbf{a}_c$ and hence can alleviate overfitting.

\subsubsection{Restriction in individual level}
At the individual level, $\mathbf{a}_u$ and $\mathbf{a}_c$ are required to generate unbiased $\mathbf{\bar{e}}_{u}$ and $\mathbf{\bar{e}}_{c}$ from the biased $\mathbf{\bar{e}}^{\prime}_{u}$ and $\mathbf{\bar{e}}^{\prime}_{c}$ , respectively. 
To meet this requirement, we introduce a user restriction loss $\mathcal{L}_{rsu}$ and a cluster restriction loss $\mathcal{L}_{rsc}$.
$\mathcal{L}_{rsu}$ measures the Euclidean distance between $\mathbf{\bar{e}}_{u}$ and $\mathbf{a}_u \odot \mathbf{\bar{e}}^{\prime}_{u}$, where $\odot$ is the element-wise product.
Similarly, $\mathcal{L}_{rsc}$ measures the Euclidean distance between $\mathbf{\bar{e}}_{c}$ and $\mathbf{a}_c \odot \mathbf{\bar{e}}^{\prime}_{c}$.
$\mathcal{L}_{rsu}$ and $\mathcal{L}_{rsc}$ can be written as:
\begin{equation}
\begin{aligned}
&\mathcal{L}_{rsu}= \frac{1}{|\mathcal{U}|}\sum_{u \in \mathcal{U}} \|\mathbf{\bar{e}}_{u} - \mathbf{a}_u \odot \mathbf{\bar{e}}^{\prime}_{u}\|_{2}^{2}, \\
&\mathcal{L}_{rsc}= \frac{1}{|\mathcal{C}|}\sum_{c \in \mathcal{C}} \|\mathbf{\bar{e}}_{c} - \mathbf{a}_c \odot \mathbf{\bar{e}}^{\prime}_{c}\|_{2}^{2}.
\end{aligned}
\end{equation} 
Minimizing $\mathcal{L}_{rsu}$ and $\mathcal{L}_{rsc}$ forces $\mathbf{a}_u$ and $\mathbf{a}_c$ to mitigate the domain bias for user $u$ and cluster $c$.
Therefore, $\mathbf{a}_u$ and $\mathbf{a}_c$ can be learned as the debiasing vectors.

\subsection{Model Optimization \label{sec: train}}
Because of removing the non-linear projection in the graph convolutional layers, the trainable parameters of our model $\theta$ are the user embedding $\mathbf{e}_u$, the parameters of the dimension reduction layer ($\mathbf{w}$ and $b$), the user debiasing embedding $\mathbf{a}_u$, and the cluster debiasing embedding $\mathbf{a}_c$.
We consider these to optimize our model.
We use the Bayesian Personalized Ranking (BPR) loss \cite{rendle2009BPR} to learn users' item preference scores.
The BPR loss is obtained as:
\begin{equation}
\begin{aligned}
\mathcal{L}_{bpr}=-\sum_{(u,v,c) \in \mathcal{R}_t} \ln \sigma\left(\hat{y}_{u v}-\hat{y}_{u v^{-}}\right),
\end{aligned}
\end{equation}
where $v^{-}$ is a negative item randomly sampled from $\mathcal{\bar{V}}_u$.

The total loss is measured by combining the dimension reduction loss $\mathcal{L}_{dr}$, the restriction loss $\mathcal{L}_{rs}$, and the BPR loss $\mathcal{L}_{bpr}$, that is,
\begin{equation}
\begin{aligned}
\mathcal{L}= \mathcal{L}_{bpr} +\lambda_{1} \mathcal{L}_{rs} + \lambda_{2} \mathcal{L}_{dr} + \lambda_{3}\|\theta\|^{2},
\end{aligned}
\end{equation}
where $\mathcal{L}_{rs} = \mathcal{L}_{rsp}+\mathcal{L}_{rsu}+\mathcal{L}_{rsc}$.
$\lambda_{1}$, $\lambda_{2}$, and $\lambda_{3}$ are hyper-parameters used to balance the weight between losses. 
We employ a gradient descent algorithm to optimize $\theta$ by minimizing $\mathcal{L}$.

\section{Experiments \label{sec:exp}}

The objective of our experiments is to answer the following research questions:
\begin{itemize}
    \item \textbf{RQ1}: How does SCDGN perform on recommendations compared with state-of-the-art methods?
    \item \textbf{RQ2}: Does the proposed debiasing learning framework benefit recommendations? 
    \item \textbf{RQ3}: Does the semantic clustering facilitate recommendations by fusing the knowledge from another domain?
    \item \textbf{RQ4}: Does the metric-invariant dimension reduction approach work in improving recommendation performance?
    \item \textbf{RQ5}: How does K (the recommendation list size) affect the recommendation accuracy of SCDGN? 
\end{itemize}

\subsection{Experiment Setting \label{sec: expset}} 
\subsubsection{Dataset.} 
We conducted experiments on two proprietary datasets and three widely used public datasets to investigate the recommendation performance of SCDGN in practical applications and for benchmarking purposes.

The public datasets contain a subset of MovieLens25M\footnote{grouplens.org/datasets/movielens/25m/} and two subsets of Amazon\footnote{jmcauley.ucsd.edu/data/amazon/}. 
The subset of MovieLens25M (ML) contains movie ratings from \textit{30/9/2016} to \textit{1/10/2018}, where the movie descriptions in ML were collected from the public API of TMDB\footnote{www.themoviedb.org/documentation/api}. 
The two subsets of Amazon include an AmazonBook (AB) dataset and an AmazonMovie (AM) dataset. 
AB and AM contain book and movie ratings from \textit{30/9/2016} to \textit{3/10/2018}, respectively, as well as textual descriptions of the books and movies.

The private datasets have an online advertisement dataset (ADs) \cite{yonekawa2019advertiser} and an e-commerce dataset (E-com).
ADs contains web browsing records from \textit{1/8/2017} to \textit{31/8/2017} on an ads platform and the textual content of Web pages.
E-com provides purchase records from an e-commerce platform and the textual descriptions of products, where the purchase records in E-com have the same period as that of ADs. 

We measured three cross-domain recommendation tasks, where each recommendation task contains an auxiliary source domain and a relatively sparse target domain.
We defined A$\rightarrow$B as a cross-domain recommendation task, where $A$ is the source domain, and $B$ is the target domain. 
The recommendation tasks include (1) ADs$\rightarrow$E-com, (2) ML$\rightarrow$AM, and (3) ML$\rightarrow$AB.
Besides, we also measured the source-target inversion version of the above tasks: (4) E-com$\rightarrow$ADs, (5) AM$\rightarrow$ML, and (6) AB$\rightarrow$ML.
For each source domain, we selected users who have $3$ to $10$ interaction records and items that have $10$ to $15$ interaction records to fit a dense setting. 
Inversely, for each target domain, we selected users who have $3$ to $5$ interactions and items that have $5$ to $15$ interactions to form a relatively sparse environment.
Some basic information about the pre-processed datasets is summarized in Table \ref{tab:dataset}.

\begin{table}[t]
\centering
\caption{Basic information on the datasets we used. \#Int./U is the average number of interactions per user.}
\begin{tabular}{c|lllll}
\hline
&Dataset        & \#Users       & \#Items     & \#Interactions        &\#Int./U      \\
\hline
            &ML         &18,232     &14,435     &421,803    &23.14         \\
As          &AM         &22,046     &7,814      &104,216    &4.73         \\
Source      &AB         &27,662     &12,708     &129,899    &4.70         \\
            &ADs        &18,829     &12,253     &360,880    &19.17     \\
            &E-com      &17,418     &6,142      &81,499     &4.68     \\
\hline
            &ML         &6,298      &9,873      &31,445     &4.99         \\
As          &AM         &8,566      &6,752      &39,696     &4.63         \\
Target      &AB         &13,350     &10,477     &61,004     &4.57         \\
            &ADs        &11,010     &12,031     &55,050     &5.00     \\
            &E-com      &12,558     &5,118      &46,871     &3.73     \\
\hline
\end{tabular}
\label{tab:dataset}
\end{table}

\subsubsection{Evaluation criteria}
For each user in target domains, we took this user's last and second-last interactions to form the test and validation sets, respectively.
The remaining interactions were used as the training set.
Then, we randomly sampled 99 items that had no interaction with this user and ranked the target item among the 100 items.
The result for the top-K recommendations was measured by the widely used \textit{Hit Ratio} (HR) and \textit{Normalized Discounted Cumulative Gain} (NDCG) \cite{jarvelin2002cumulated}.

\begin{table*}[t]
\centering
\caption{Comparison between our proposal and state-of-the-art by using HR@K and NDCG@K. Performances ± 95\% confidence intervals are reported. Bold shows the winner.}
\begin{tabular}{l|l|lll|lll}
\hline
&                   & &ADs $\rightarrow$ E-com &                       & &E-com $\rightarrow$ ADs & \\ \cline{2-8}
                    &Method         &HR@1           &HR@5             &NDCG@5         &HR@1           &HR@5             &NDCG@5 \\ \hline
                    &NeuCF          &0.323±0.024    &0.442 ± 0.025    &0.348 ± 0.022  &0.058 ± 0.008  &0.136 ± 0.019    &0.080 ± 0.010  \\ 
singe-domain RS     &LightGCN       &0.368±0.005    &0.435 ± 0.005    &0.384 ± 0.006  &0.264 ± 0.005  &0.352 ± 0.006    &0.282 ± 0.005  \\
                    &LightGCN (wt)    &\textbf{0.383 ± 0.007}    &0.466 ± 0.005  &0.400 ± 0.008  &\textbf{0.322 ± 0.007}  &0.472 ± 0.006    &\textbf{0.354 ± 0.006}   \\ \cline{1-8}

                    &$s^2$-Meta (wt)  &0.282 ± 0.032    &0.372 ± 0.011    &0.337 ± 0.020  &0.033 ± 0.009  &0.106 ± 0.022  &0.067 ± 0.013\\ 
cross-domain RS     &RecSys-DAN (wt)  &0.254 ± 0.016    &0.352 ± 0.019    &0.277 ± 0.016  &0.043 ± 0.006  &0.114 ± 0.010  &0.065 ± 0.007  \\
                    &ESAM (wt)        &0.355 ± 0.020    &0.459 ± 0.017    &0.378 ± 0.019  &0.200 ± 0.013  &0.379 ± 0.014  &0.248 ± 0.012 \\
                    &CFAA (wt)        &0.359 ± 0.018    &0.485 ± 0.019    &0.387 ± 0.020  &0.128 ± 0.009  &0.300 ± 0.015  &0.181 ± 0.010 \\\cline{2-8}
                    &SCDGN (ours)   &0.380 ± 0.014    &\textbf{0.496 ± 0.012}   &\textbf{0.410 ± 0.012}  &0.312 ± 0.006    &\textbf{0.489 ± 0.005}   &\textbf{0.354 ± 0.005}  \\ \hline

&                   & &ML $\rightarrow$ AM &                       & &AM $\rightarrow$ ML & \\ \cline{2-8}
                    &Method         &HR@1           &HR@5             &NDCG@5         &HR@1           &HR@5             &NDCG@5 \\ \hline
                    &NeuCF          &0.075 ± 0.021    &0.154 ± 0.013    &0.097 ± 0.016  &0.074 ± 0.022  &0.198 ± 0.024    &0.113 ± 0.018    \\ 
singe-domain RS     &LightGCN       &0.160 ± 0.003    &0.218 ± 0.002    &0.175 ± 0.002  &0.139 ± 0.009  &0.266 ± 0.011    &0.174 ± 0.009    \\
                    &LightGCN (wt)    &0.168 ± 0.002    &0.244 ± 0.003    &0.189 ± 0.002  &0.145 ± 0.014  &0.300 ± 0.018    &0.189 ± 0.013    \\ \cline{1-8}

                    &$s^2$-Meta (wt)  &0.059 ± 0.002    &0.126 ± 0.001    &0.094 ± 0.001  &0.027 ± 0.001  &0.103 ± 0.001    &0.063 ± 0.001  \\
cross-domain RS     &RecSys-DAN (wt)  &0.108 ± 0.006    &0.163 ± 0.007    &0.124 ± 0.006  &0.045 ± 0.008  &0.111 ± 0.010    &0.066 ± 0.008    \\
                    &ESAM (wt)        &0.095 ± 0.016    &0.169 ± 0.010    &0.115 ± 0.012  &0.144 ± 0.017  &0.301 ± 0.021    &0.189 ± 0.017 \\
                    &CFAA (wt)        &0.111 ± 0.010    &0.189 ± 0.008    &0.133 ± 0.008  &0.124 ± 0.015  &0.274 ± 0.022    &0.168 ± 0.016  \\\cline{2-8}
                    &SCDGN (ours)   &\textbf{0.181 ± 0.005}    &\textbf{0.260 ± 0.006}   &\textbf{0.200 ± 0.005}  &\textbf{0.180 ± 0.012}    &\textbf{0.356 ± 0.013}   &\textbf{0.229 ± 0.011}  \\ \hline  
&                   & &ML $\rightarrow$ AB &                       & &AB $\rightarrow$ ML & \\ \cline{2-8}
                    &Method         &HR@1           &HR@5             &NDCG@5         &HR@1           &HR@5             &NDCG@5 \\ \hline
                    &NeuCF          &0.091 ± 0.011    &0.186 ± 0.022    &0.116 ± 0.013  &0.074 ± 0.022  &0.198 ± 0.024    &0.113 ± 0.018    \\ 
singe-domain RS     &LightGCN       &0.175 ± 0.006    &0.269 ± 0.005    &0.197 ± 0.005  &0.139 ± 0.009  &0.266 ± 0.011    &0.174 ± 0.009    \\
                    &LightGCN (wt)    &0.173 ± 0.007    &0.278 ± 0.006    &0.199 ± 0.006  &0.145 ± 0.014  &0.300 ± 0.018    &0.189 ± 0.013    \\ \cline{1-8}

                    &$s^2$-Meta (wt)  &0.056 ± 0.001    &0.170 ± 0.002    &0.114 ± 0.001  &0.025 ± 0.001  &0.103 ± 0.001    &0.063 ± 0.001    \\
cross-domain RS     &RecSys-DAN (wt)  &0.067 ± 0.005    &0.136 ± 0.009    &0.087 ± 0.007  &0.043 ± 0.006  &0.114 ± 0.010    &0.065 ± 0.007    \\
                    &ESAM (wt)        &0.099 ± 0.019    &0.212 ± 0.019    &0.129 ± 0.015  &0.138 ± 0.019  &0.299 ± 0.022    &0.185 ± 0.018  \\
                    &CFAA (wt)        &0.111 ± 0.016    &0.228 ± 0.016    &0.141 ± 0.014  &0.132 ± 0.015  &0.292 ± 0.024    &0.179 ± 0.017 \\\cline{2-8}
                    &SCDGN (ours)   &\textbf{0.199 ± 0.011}    &\textbf{0.321 ± 0.008}   &\textbf{0.228 ± 0.010}  &\textbf{0.181 ± 0.011}    &\textbf{0.350 ± 0.013}   &\textbf{0.227 ± 0.010}  \\    

\hline
\end{tabular}
\label{tab:comprslt}
\end{table*}

\subsubsection{Evaluated methods} 
To measure the validity of the semantic information coming from the source data, we compared our method with the following state-of-the-art methods:
\paragraph{Single-domain recommendations (SDRs)}
\begin{itemize}
\item{\textbf{NeuCF} \cite{he2017neural}} jointly learns a neural network and a matrix factorization model. 
\item{\textbf{LightGCN} \cite{he2020lightgcn}} is a light graph convolutional network that enhances the user and item embeddings with the learned structural information from the user-item interaction graph.
\end{itemize}
\paragraph{Cross-domain recommendations (CDRs)}
\begin{itemize}
\item{\textbf{$s^2$-Meta} \cite{du2019s2meta}} develops a meta-learning framework to generate individual models for different scenarios, where scenarios are denoted as domains. 
We represented users by the average of their interacted items to run $s^2$-Meta between domains with no overlapping users and items.
\item{\textbf{RecSys-DAN} \cite{wang2020rsdan}} trains a source user preference predicting model via the source domain data and then transfers the learned user preference patterns by aligning user preference patterns between source and target models.
\item{\textbf{ESAM} \cite{Zhihong2020ESAM}} adopts attribute correlation alignment to improve long-tail recommendation performance by suppressing inconsistent distribution between items from source and target domains.
\item{\textbf{CFAA} \cite{liu2022CFAA}} proposes an embedding attribution alignment module to reduce the discrepancy of attribution distributions and relations between source and target domains. 
\end{itemize}

For fair comparisons, we aligned the base model for all cross-domain methods with LightGCN, where this base model is equal to our SCDGN without the cross-domain user-cluster graph part. 
Besides, we replaced the randomly initialized item embedding with our pre-trained semantic item embedding in Section \ref{sec:embed} for all cross-domain comparisons and LightGCN, where these methods are identified with (wt).

\subsubsection{Implementation details}
The codes of NeuCF\footnote{github.com/yihong-chen/neural-collaborative-filtering}, LightGCN\footnote{github.com/gusye1234/LightGCN-PyTorch}, and $s^2$-Meta\footnote{github.com/THUDM/ScenarioMeta} were obtained from the corresponding GitHub repositories. 
Our SCDGB, ESAM, and CFAA were implemented by using PyTorch framework and can be found in a GitHub repository\footnote{github.com/ZL6298/SCDGN}. 
We used Adam to optimize the model parameters and speed up the training process with the mini-batch trick. 
For hyper-parameters, the learning rate was 0.001 for the recommendation tasks on private datasets and 0.01 for the cases on public datasets.
The cluster number was 200.
The embedding size of $\mathbf{e}_u$ was 32. 
The mini-batch size was 1024.
The restriction loss balance factor $\lambda_1$ was set to 1, 0.001, and 0.0001 for the recommendation task on ML$\rightarrow$AM, E-com$\rightarrow$ADs, and ADs$\rightarrow$E-com, respectively. 
$\lambda_1$ was set to 0.01 for the recommendation task on AM$\rightarrow$ML, ML$\rightarrow$AB, and AB$\rightarrow$ML.
The dimension reduction loss balance factor $\lambda_2$ was set to 1 for the recommendation task on private datasets, ML$\rightarrow$AM, and ML$\rightarrow$AB, where it was set to 10 for the recommendations on AM$\rightarrow$ML and AB$\rightarrow$ML. 
The weight of the regularization term $\lambda_3$ was set to 0.01 for the recommendation task on private datasets and 0.1 for the case on public datasets.
The user-cluster graph convolutional layer number $P$ was set to 2 for recommendations on public datasets and 1 for the private datasets.
For fair comparisons, we set the same user-item graph convolutional layer number $Q$ as 3 for all comparisons except NeuCF.
All these hyper-parameters were tuned on the validation set.

\subsection{Performance Comparison (RQ1) \label{sec: perform}}
We report the average recommendation performances on the test set of each target domain.
The comparison results are listed in Table \ref{tab:comprslt}.
This table shows that SCDGN outperforms other competitors on HR@5, NDCG@5, and HR@1 (in most cases). 
Besides, SCDGN achieves a remarkable improvement on four public recommendation tasks. 
This observation empirically demonstrates that our SCDGN effectively leverages the semantic information on the source domains to improve the recommendations in the target domains.
For single-domain RSs, we find that LightGCN (wt) achieves the best performance, LightGCN the second best, and NeuCF the worst.
This is because the target semantic information and the graph convolutional network yield a better performance. 
For cross-domain RSs, although they transfer interaction patterns or align embedding space from the source domain to the target domain, they perform worse than the single-domain method, i.e., LightGCN, in most cases. 
This result indicates that domain bias in interaction patterns causes the negative transfer issue and an inferior performance.

\subsection{Vs. CDRS with Overlapping Users (RQ1)} \label{sec:compwithRSreqoveruser}
To further investigate the effectiveness of the proposed method, we identified overlapping users between AM and AB and conducted experiments to compare our SCDGN with CDRSs that require overlapping users.
Some basic information of the datasets used in this experiment is summarized in Table \ref{tab:overlappingdataset}.

\begin{table}[!t]
\centering
\caption{Basic information on the datasets with only the overlapping users. \#Int./U is the average number of interactions per user.}
\begin{tabular}{c|lllll}
\hline
&Dataset        & \#Users       & \#Items     & \#Interactions        & \#Int./U      \\
\hline         
As          &AM       &1,315     &5,458     &15,169    &11.54    \\
Source      &AB      &722     &2,894     &6,485     &8.99    \\
\hline
As          &AM     &722     &3,337     &5,870     &8.13     \\
Target      &AB     &1,315     &4,246     &7,458     &5.67     \\
\hline
\end{tabular}
\label{tab:overlappingdataset}
\end{table}

\begin{table}[t]
\centering
\caption{Comparison between our proposal and CDRSs that require overlapping users}
\begin{tabular}{l|lll}
\hline
Scenario            &Method       &HR@1           &HR@5                 \\ 
\hline
AM$\rightarrow$    &CGN          &0.036 ± 0.001  &0.131 ± 0.002       \\   
AB               &BiTGCF       &0.059 ± 0.002  &\textbf{0.217 ± 0.002}        \\ \cline{2-4}
                    &SCDGN (ours)        &\textbf{0.094 ± 0.003}  &0.171 ± 0.003 \\
\hline
AB$\rightarrow$  &CGN          &0.022 ± 0.001  &0.173 ± 0.006       \\   
AM                 &BiTGCF       &0.087 ± 0.003  &\textbf{0.266 ± 0.003}         \\ \cline{2-4}
                    &SCDGN (ours)        &\textbf{0.127 ± 0.002}  &0.209 ± 0.003 \\
\hline
\end{tabular}
\label{tab:results_bridge}
\end{table} 

\subsubsection{Evaluated methods}
We compared our method with the following state-of-the-art CDRSs: 
\begin{itemize}
\item{\textbf{CGN} \cite{zhang2020itemsetmap}} develops generative models for each domain to produce users' interacted itemset.
Then, the domain knowledge is transferred via mapping the generated users' interacted itemset between domains.  
\item{\textbf{BiTGCF} \cite{liu2020transgcf}} is a GNN-based CDRS that learns domain-specific feature propagation layers to alleviate the domain bias.
It transfers knowledge in individual level by matching the overlapping users from both domains. 
\end{itemize}

\subsubsection{Comparison results}
Table \ref{tab:results_bridge} shows the comparison results on HR@1 and HR@5. 
We observe that our SCDGN remarkably outperforms CGN and achieves a competitive performance with BiTGCF.
CGN transfers users' interaction patterns between domains and neglects the domain bias in user preference patterns. 
As a result, it yields a degraded performance.
Both BiTGCF and SCDGN propose approaches to alleviating the impact of the bias and thus outperform CGN. 
In addition, it is worth mentioning that SCDGN involves no user assignment, suggesting that SCDGN has a broader application than CGN and BiTGCF.

\begin{figure*}[t]
    \begin{center}
        \subfigure[ML $\rightarrow$ AM]{
        \includegraphics[width=0.235\linewidth]{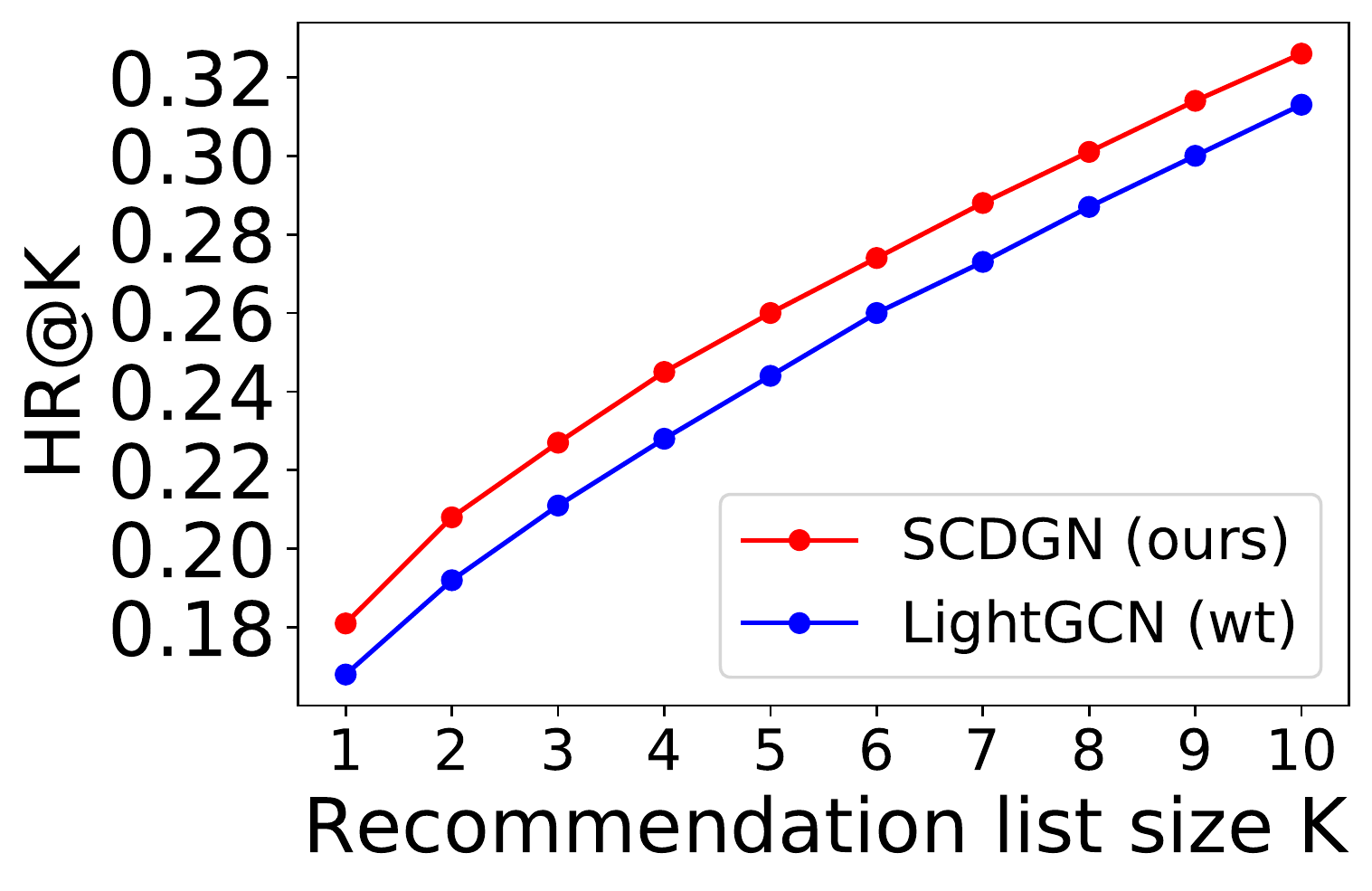}}
        \subfigure[AM $\rightarrow$ ML]{
        \includegraphics[width=0.235\linewidth]{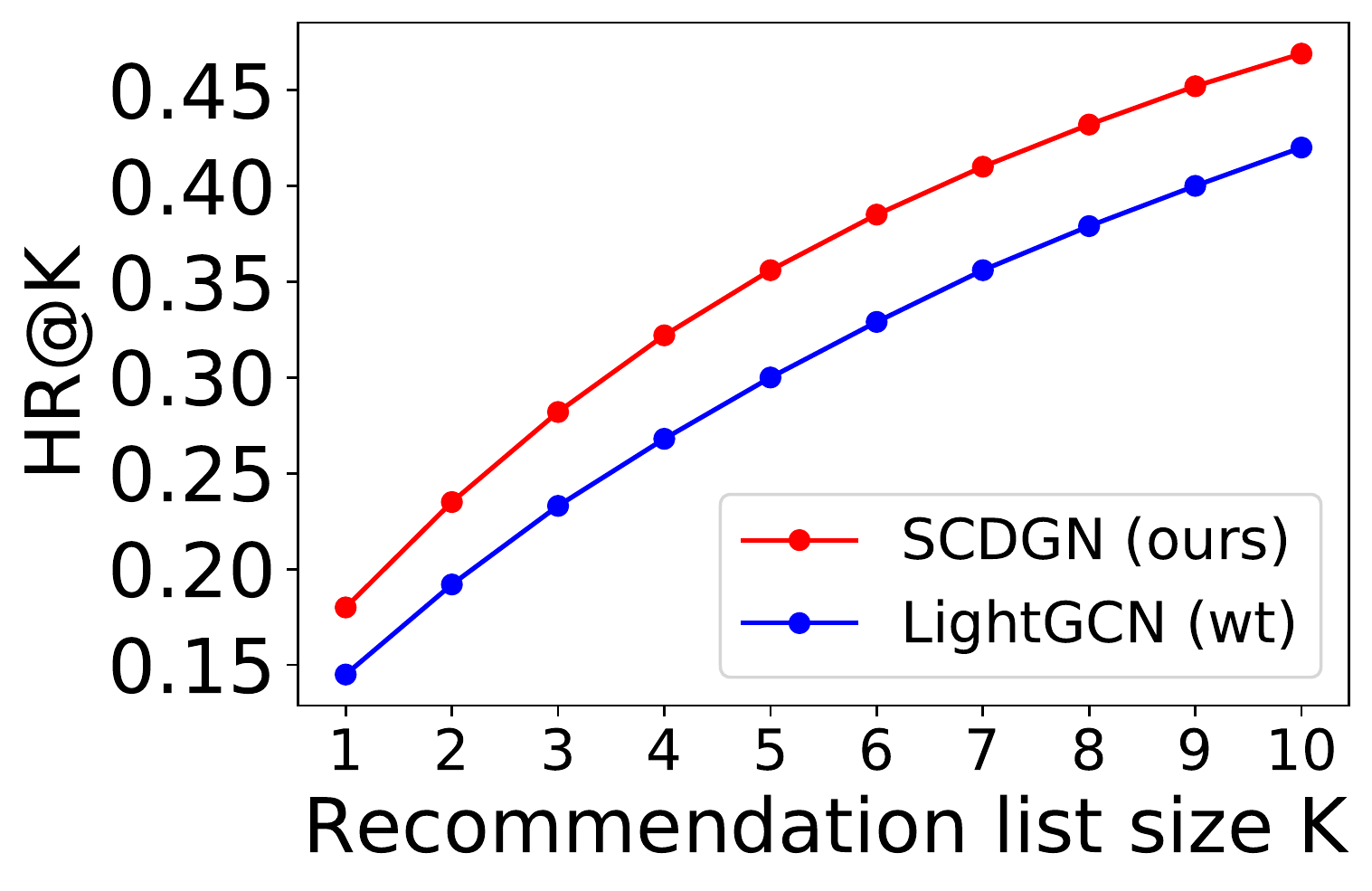}}
        \subfigure[ML $\rightarrow$ AB]{
        \includegraphics[width=0.235\linewidth]{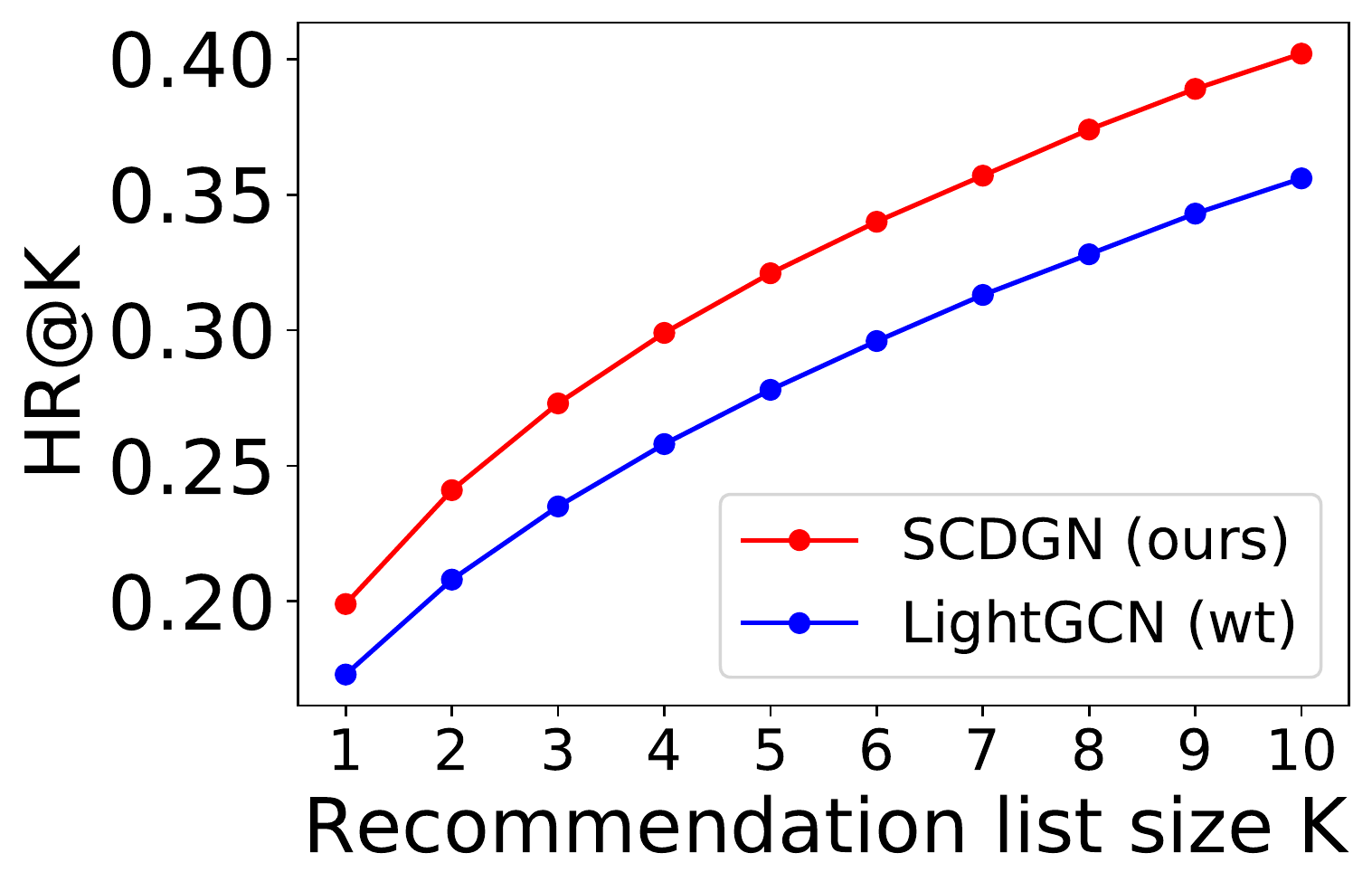}}
        \subfigure[AB $\rightarrow$ ML]{
        \includegraphics[width=0.235\linewidth]{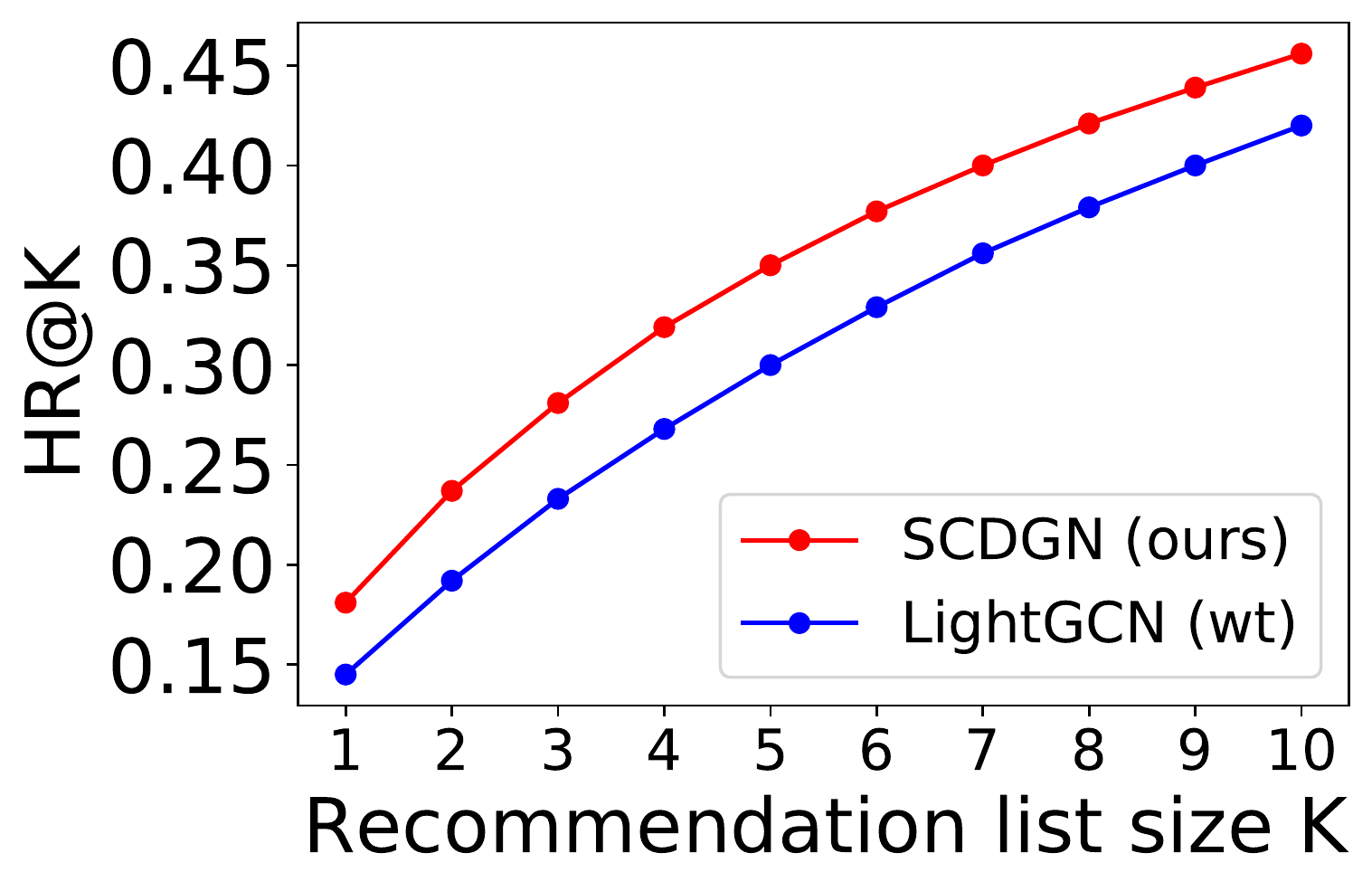}}
        \subfigure[ML $\rightarrow$ AM]{
        \includegraphics[width=0.235\linewidth]{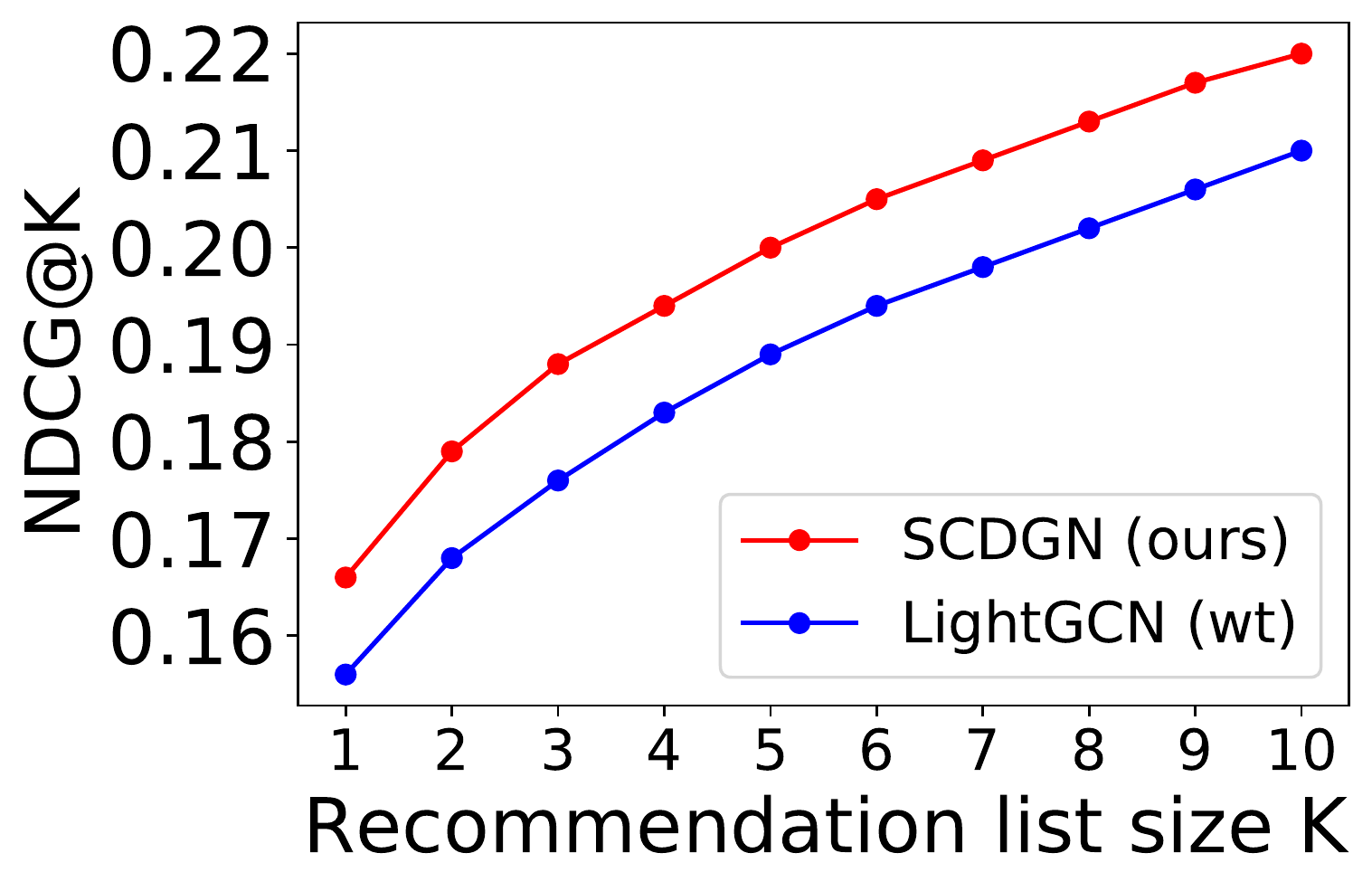}}
        \subfigure[AM $\rightarrow$ ML]{
        \includegraphics[width=0.235\linewidth]{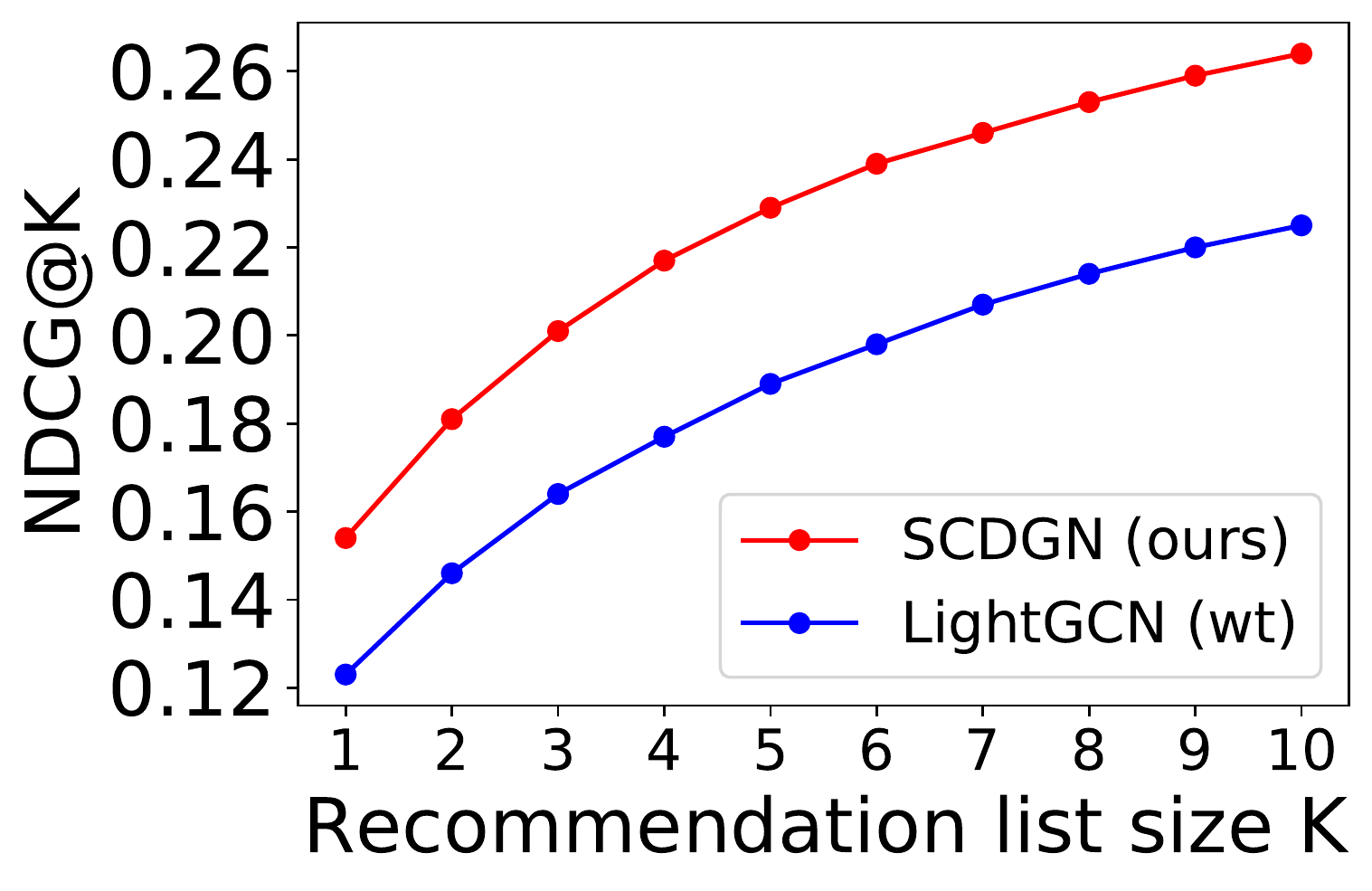}}
        \subfigure[ML $\rightarrow$ AB]{
        \includegraphics[width=0.235\linewidth]{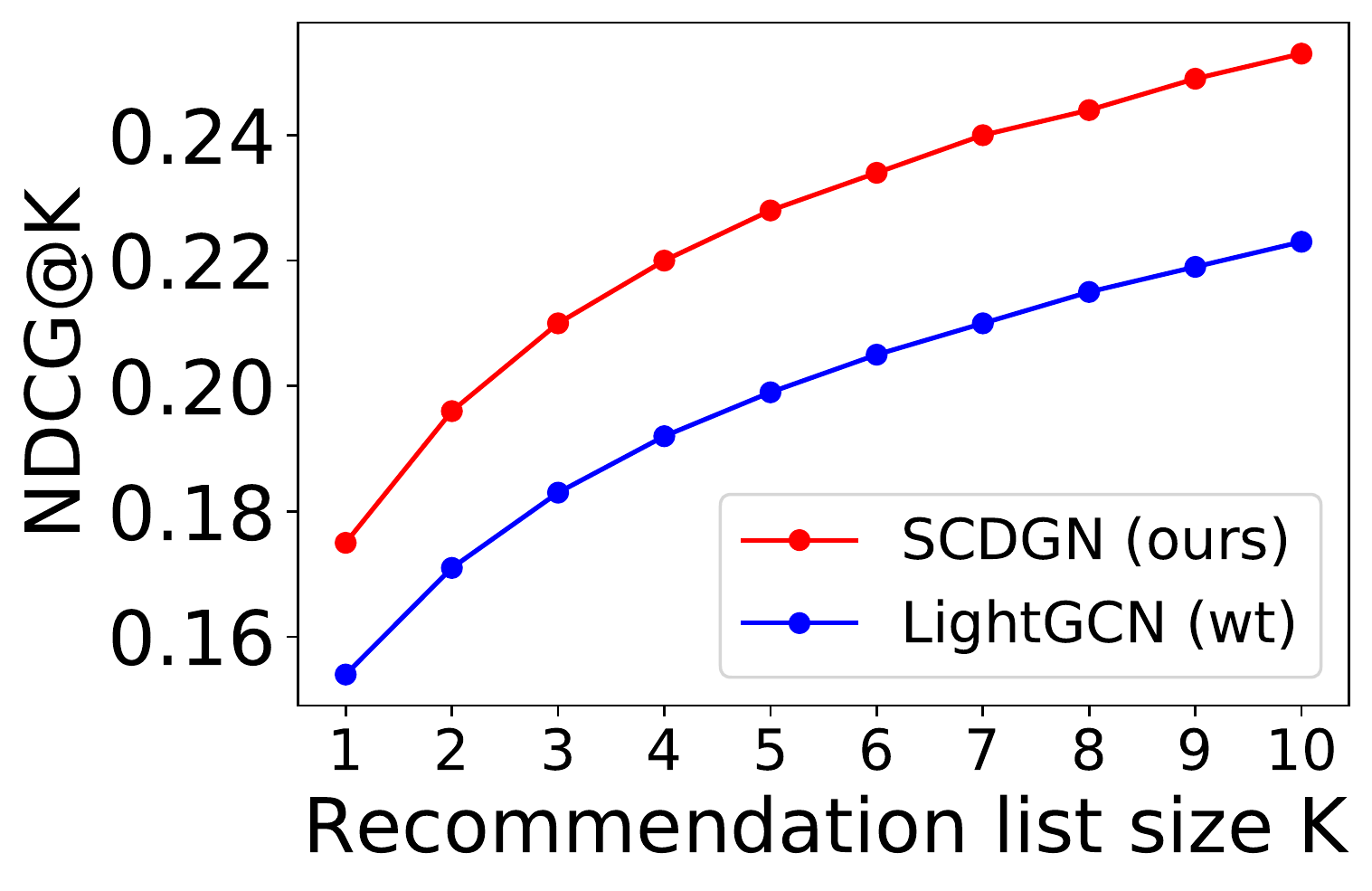}}
        \subfigure[AB $\rightarrow$ ML]{
        \includegraphics[width=0.235\linewidth]{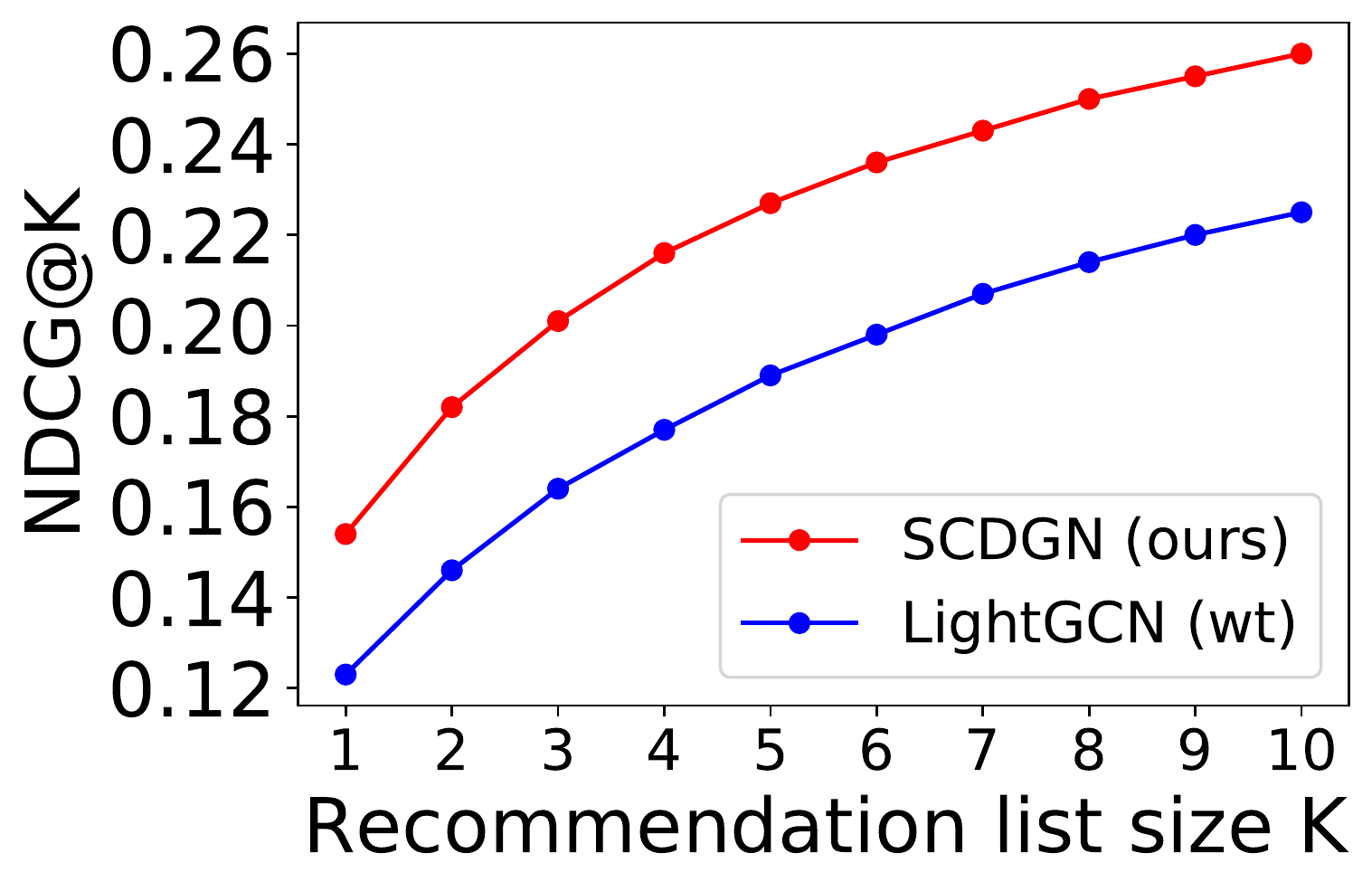}}     
    \end{center}
    \caption{Impact of K}
    \label{fig:ImpactOfK}
\end{figure*}

\begin{table}[t]
\centering
\caption{Performances of variants of SCDGN}
\begin{tabular}{l|lll}
\hline
    Dataset      &Method      &HR@5           &NDCG@5         \\ \hline
                         &w/o SI        &0.244 ± 0.003 &0.189 ± 0.002 \\                        
    ML $\rightarrow$     &w/o DRloss    &0.240 ± 0.003  &0.186 ± 0.002 \\ 
    AM                   &w/o DB        &0.229 ± 0.003 &0.177 ± 0.002 \\ \cline{2-4}
                         &SCDGN         &\textbf{0.260 ± 0.006} &\textbf{0.200 ± 0.005} \\ \hline
                         &w/o SI        &0.278 ± 0.006 &0.199 ± 0.006 \\                        
    ML $\rightarrow$     &w/o DRloss    &0.314 ± 0.008 &0.223 ± 0.010 \\ 
    AB                   &w/o DB        &0.253 ± 0.006 &0.183 ± 0.007 \\ \cline{2-4}
                         &SCDGN         &\textbf{0.321 ± 0.008} &\textbf{0.228 ± 0.01} \\ 
\hline
\end{tabular}
\label{tab:abrslt}
\end{table}

\subsection{Ablation Study (RQ2 \& RQ3 \& RQ4) \label{sec: abstudy}}
To study the impact of different components of SCDGN, we conducted ablation studies on ML$\rightarrow$AM and ML$\rightarrow$AB with some variants of SCDGN, including (1) w/o SI: SCDGN without user-cluster graph information, which is equal to LightGCN (wt), (2) w/o DRloss: SCDGN without the dimension reduction loss $\mathcal{L}_{dr}$, and (3) w/o DB: SCDGN without debiasing learning mechanism.
Table \ref{tab:abrslt} shows HR@5 and NDCG@5 of SCDGN and its variants.
From this table, we can see that all the information from the user-cluster graph, the metric-invariant dimension reduction, and the debiasing learning boost recommendation accuracy.
Specifically, the results decrease the most without the proposed debiasing learning approach.
This observation demonstrates that it is necessary to handle domain bias even when transferring the semantic cluster-level interaction information.
Besides, the decrement of results on w/o DRloss indicates the effectiveness of constraining the metric relationship when reducing dimension in a sparse domain.

\subsection{Impact of Recommendation List Size (RQ5) \label{sec:impactOfK}}
To investigate the impact of the recommendation list size K, we conducted experiments on public datasets by varying K.
We used LightGCN (wt) as a competitor, as it is the best baseline.
Figure \ref{fig:ImpactOfK} shows the results on HR@K and NDCG@K.
From these figures, we can see that SCDGN outperforms LightGCN (wt) consistently.
This finding indicates that SCDGN successfully extracts the unbiased structural knowledge from the cross-domain cluster-level graph, where this knowledge is effective in producing a better recommendation.
For the cross-domain recommendations between ML and AB, our SCDGN achieves a greater improvement than LightGCN (wt) over both HR@K and NDCG@K. 
This result demonstrates that the debasing learning mechanism in SCDGN facilitates cross-domain recommendations, especially for domains with different user behaviors. 

\begin{figure}[t]
    \begin{center}
        \subfigure{
        \includegraphics[width=0.460\linewidth]{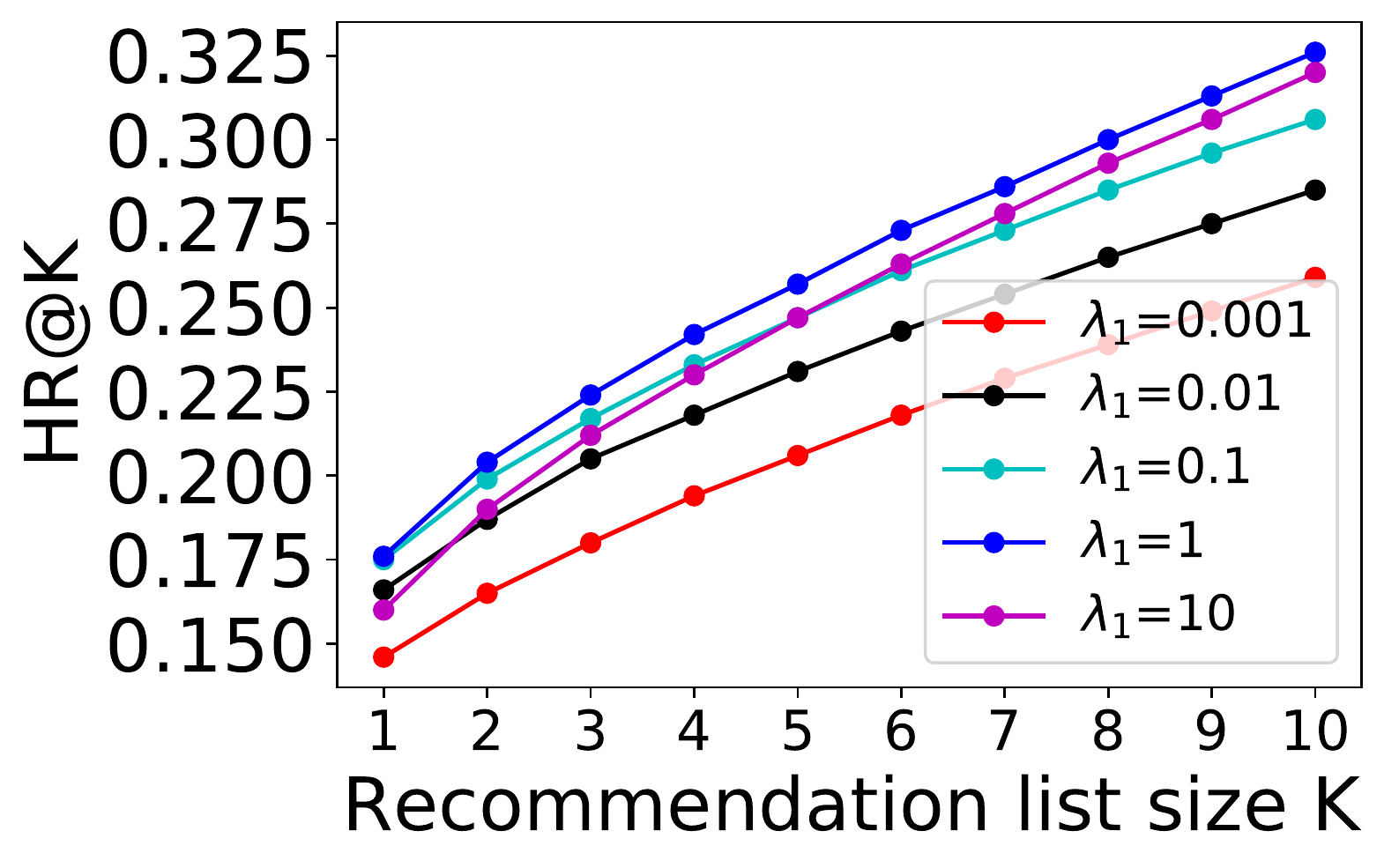}
        }
        \subfigure{
        \includegraphics[width=0.460\linewidth]{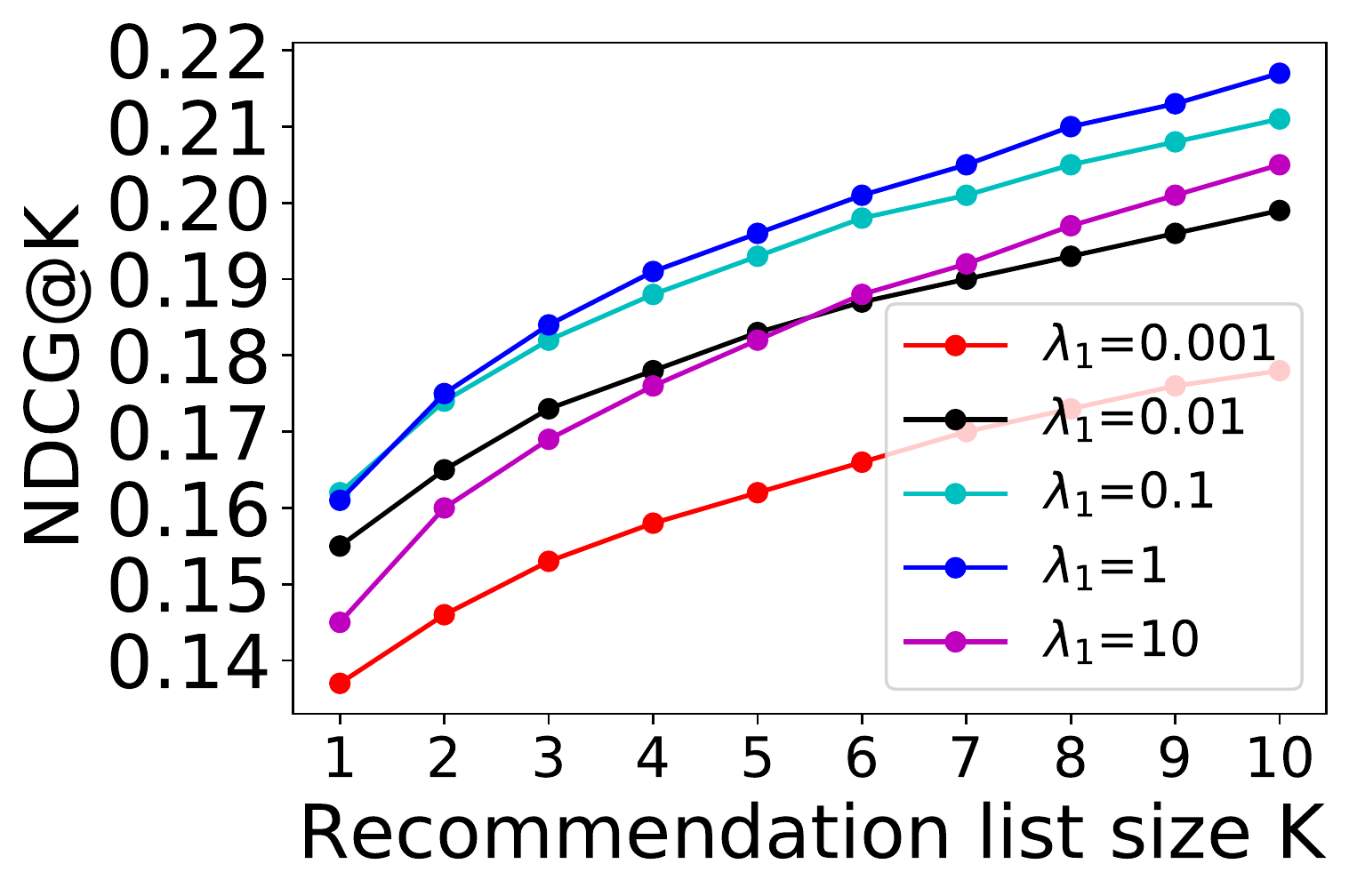}
        }
        \subfigure{
        \includegraphics[width=0.460\linewidth]{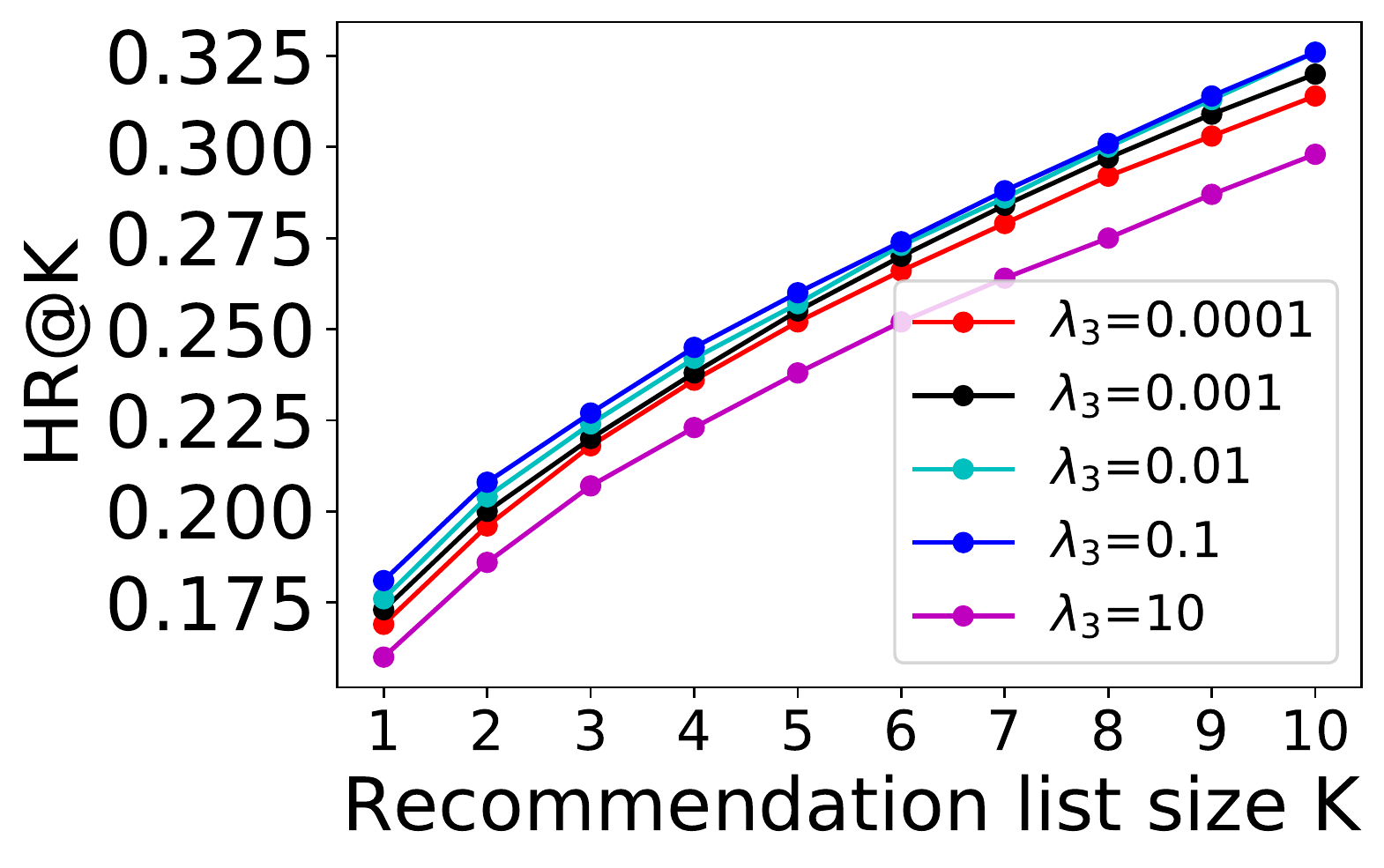}
        }
        \subfigure{
        \includegraphics[width=0.460\linewidth]{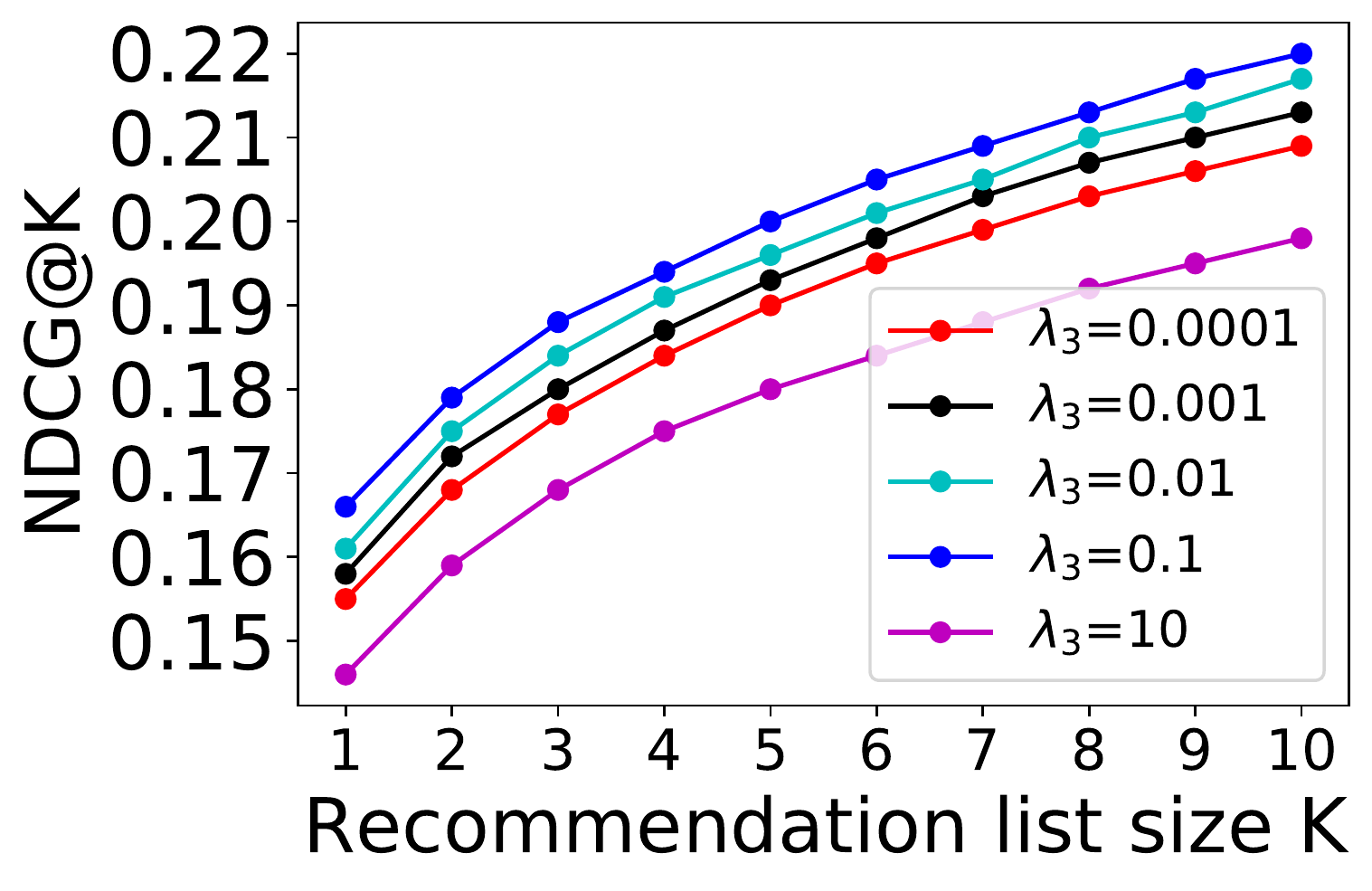}
        }
    \end{center}
\caption{Impact of $\lambda_1$ and $\lambda_3$}
\label{fig:ImpactOflambda}
\end{figure}

\subsection{Impact of Loss Balance Factors $\lambda_1$ and $\lambda_3$ \label{sec:impactOflambda}}
In this part, we conducted experiments on ML$\rightarrow$AM to discuss the impact of the hyper-parameter $\lambda_1$ and $\lambda_3$, where $\lambda_1$ and $\lambda_3$ are the factors to balance the restriction loss and the regularization term, respectively.
Figure \ref{fig:ImpactOflambda} reports the results on HR@K and NDCG@K with varying $\lambda_1$ and $\lambda_3$. 
From this figure, we found that SCDGN achieves the best performance when $\lambda_1 = 1$ and $\lambda_3 = 0.001$. 
A small $\lambda_1$ produces an under-fitting issue when learning user and item debiasing vector, resulting in an inferior performance. 
Inversely, a large $\lambda_1$ may introduce noise information from the source domain to mislead the user preference prediction of the target domain.
Besides, a proper $\lambda_3$ is necessary to prevent the optimization of SCDGN from over-fitting and under-fitting issues.

\subsection{Impact of Hyper-parameter $P$}
\begin{figure}[t]
    \begin{center}
        \subfigure{
        \includegraphics[width=0.460\linewidth]{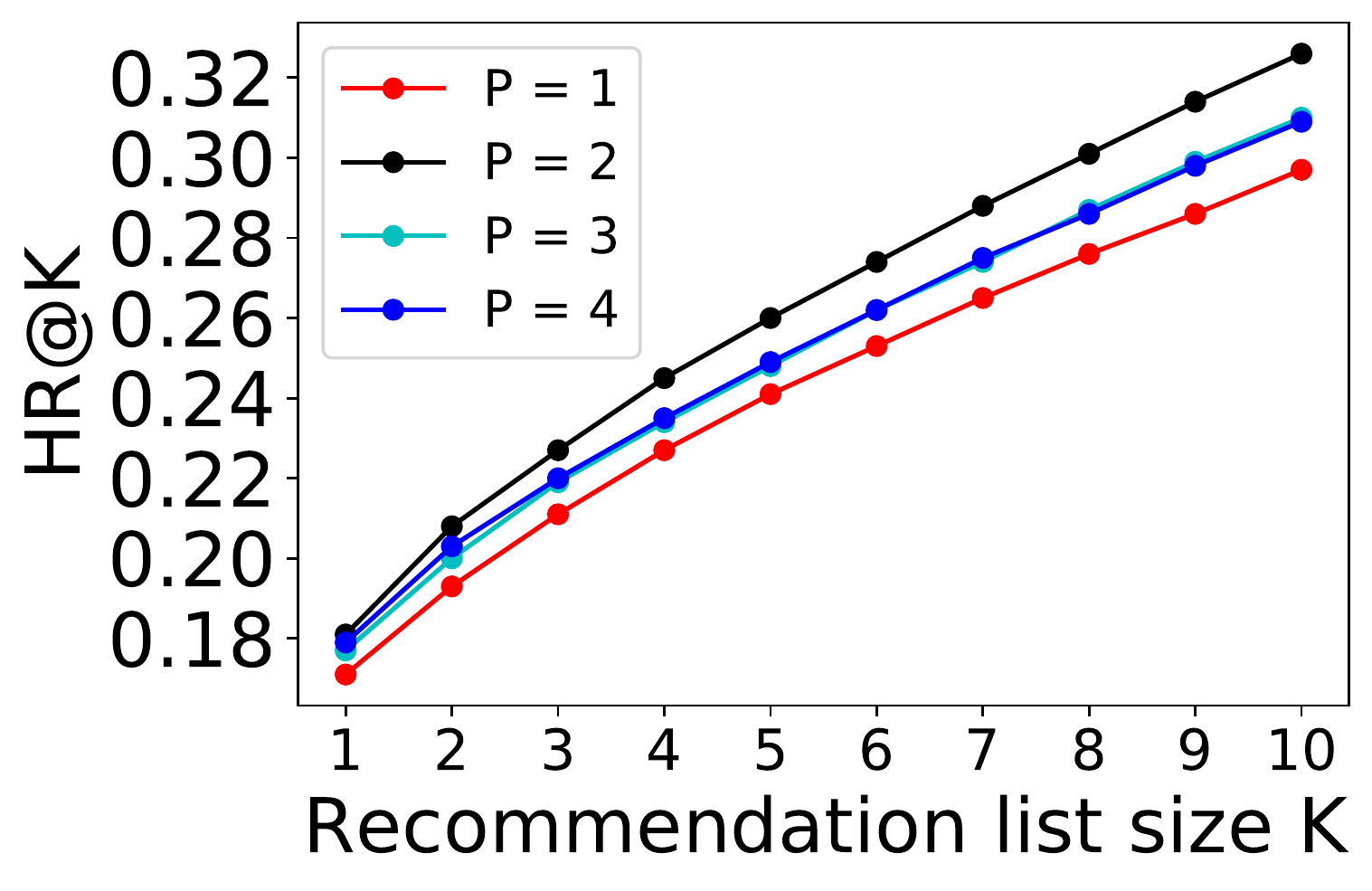}
        }
        \subfigure{
        \includegraphics[width=0.460\linewidth]{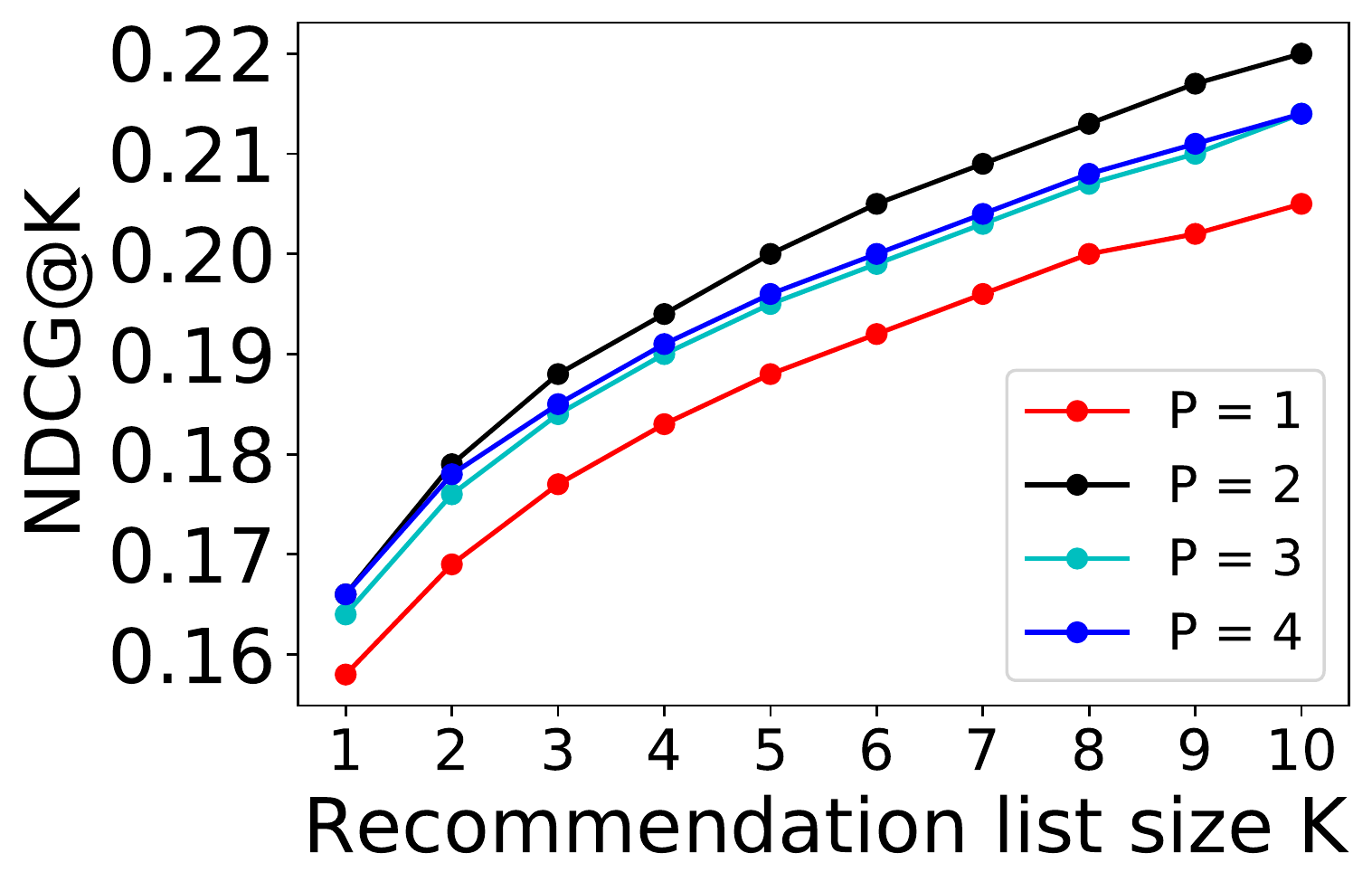}
        }
    \end{center}
\caption{Impact of $P$}
\label{fig:ImpactOfP}
\end{figure}

We conducted experiments on ML$\rightarrow$AM to empirically investigate the impact of $P$, the number of the debiasing graph convolutional layer for the cross-domain user-cluster graph.
Figure \ref{fig:ImpactOfP} shows the results of SCDGN with varying $P$ in a set \{1, 2, 3, 4\}.
From this result, we can see that the performance reaches its peak when $P=2$. 
This result indicates that a two-hop connected sub-graph can provide the best structural information to boost recommendations.

\section{Conclusion \label{sec:clus}}
In this work, we proposed a novel semantic clustering enhanced debiasing graph neural recommender system (SCDGN) for cross-domain recommendations with no overlapping user and item between source and target domains. 
SCDGN exploits semantic features as transferable knowledge to bridge domains and enrich the interaction information of the sparse target domain.
Specifically, SCDGN constructs a cross-domain user-cluster graph and develops a new debiasing graph convolutional layer to extract unbiased graph knowledge from the source domain.
SCDGN also introduces restriction losses to learn user and item debiasing vectors.
Furthermore, we developed a metric-invariant dimension reduction approach to alleviate over-fitting caused by the sparse data. 
The experimental results on public datasets and a pair of proprietary datasets demonstrate the superiority of SCDGN.

\section*{Acknowledgments}
This research is partially supported by JST CREST Grant Number JPMJCR21F2.

\bibliographystyle{IEEEtran}
\bibliography{reference}

\end{document}